\documentclass[aps, prc, twocolumn, nofootinbib, superscriptaddress, longbibliography]{revtex4-2}

\usepackage{amssymb}
\usepackage{amsthm}
\usepackage{amsmath}
\usepackage{hyperref}
\usepackage{bm}
\usepackage{qcircuit}
\usepackage{tabularx}
\usepackage{color,soul}
\usepackage[table, dvipsnames]{xcolor}
\usepackage{physics}
\usepackage{braket}
\usepackage{bbold}  
\usepackage{orcidlink}

\usepackage[caption=false]{subfig} 

\usepackage{soul}
\usepackage[normalem]{ulem}

\usepackage{todonotes}

\def\R{\mathbb{R}}

\def\Hnuc{H_{\mathrm{nuc}}}
\def\nnuc{n_{\mathrm{nuc}}}
\def\npro{n_{\mathrm{pro}}}
\def\nneu{n_{\mathrm{neu}}}
\def\norb{n_{\mathrm{q}}}

\def\psigs{\psi_{\mathrm{gs}}}
\def\svert{\raisebox{0.34ex}{\scalebox{1}[0.6]{$|$}}}
\def\dsvert{\raisebox{0.34ex}{\scalebox{1}[0.6]{$||$}}}

\definecolor{Bblue}{HTML}{1F3864}

\begin{document}

\author{Andrey~Kardashin${}^{\orcidlink{0000-0002-7966-9350}}$}
\email{andrei.kardashin@dipc.org} 
\affiliation{Donostia International Physics Center, Donostia-San Sebastián, Spain}

\author{Lewis~W.~Anderson${}^{\orcidlink{0000-0003-0269-3237}}$}
\affiliation{IBM Research, Winchester, SO21 2JN, United Kingdom}

\author{Declan~Millar${}^{\orcidlink{0000-0003-3713-8997}}$}
\affiliation{IBM Research, Winchester, SO21 2JN, United Kingdom}

\author{Ivano~Tavernelli${}^{\orcidlink{0000-0001-5690-1981}}$}
\affiliation{IBM Research, S\"aumerstrasse 4, 8803 R\"uschlikon, Switzerland}

\author{Vincent~R.~Pascuzzi${}^{\orcidlink{0000-0003-3167-8773}}$}
\affiliation{IBM Research, Yorktown Heights, NY 10598, USA}

\author{Beatriz~Romeo${}^{\orcidlink{0000-0001-8896-4565}}$}
\affiliation{Donostia International Physics Center, Donostia-San Sebastián, Spain}

\author{Francesc~Monrabal${}^{\orcidlink{0000-0002-4047-5620}}$}
\author{Stefano~Roberto~Soleti${}^{\orcidlink{0000-0002-5526-1414}}$}
\affiliation{Donostia International Physics Center, Donostia-San Sebastián, Spain}
\affiliation{Ikerbasque (Basque Foundation for Science), Bilbao, Spain}

\title{Sample based Krylov diagonalization of the nuclear shell model on a superconducting quantum computer}

\begin{abstract}

    Among the most natural applications of quantum computers is the simulation of many-body quantum systems such as atomic nuclei.
    In this paper, we compute the ground state of nuclear Hamiltonians within the nuclear shell model by applying SqDRIFT, a stochastic sample-based quantum diagonalization algorithm which uses a quantum computer to identify the most important subspace of fermionic configurations.
    We execute SqDRIFT on the \texttt{ibm\_basquecountry} superconducting quantum processor for the isotopes $^{48}$Ca, $^{28}$Si, and $^{60}$Zn with up to 44 qubits.
    To repair the noisy samples obtained from the quantum computer, we develop a configuration recovery algorithm which respects the symmetries of nuclear Hamiltonians.
    Our work shows how SqDRIFT, when combined with classical post-processing, can be used to compute accurate ground state energies of nuclear Hamiltonians in agreement with exact diagonalization in the corresponding model spaces.

\end{abstract}

\maketitle

\section{Introduction}

    Quantum computers offer powerful approaches to studying many-body systems \cite{ayral2023quantum} such as lattice models \cite{kokail2019self}, molecules \cite{mcardle2020quantum, motta2022emerging}, and nuclei \cite{holland2020optimal, savage2024quantum}.
    Beyond their intrinsic interest, accurate nuclear many-body calculations provide critical input for precision tests of the Standard Model and searches for physics beyond it, including neutrinoless double-beta decay ($0\nu\beta\beta$), electric dipole moments, superallowed $\beta$-decay, and other precision tests of fundamental symmetries~\cite{Engel13EDMs,Hardy2020,2023fundamentalsymmetriesneutronsneutrinos}.
    In particular, the observation of $0\nu\beta\beta$ would provide experimental evidence for the Majorana nature of neutrinos and give access to the absolute neutrino mass scale~\cite{dolinski2019neutrinoless, agostini2023toward}. 
    
    However, the predicted nuclear matrix elements (NMEs), the transition amplitudes governing the $0\nu\beta\beta$ decay rate, currently differ by more than a factor of two across many-body methods~\cite{engel2017status}. 
    Since the decay rate scales as the square of the NME, this factor-of-two spread translates into a significant uncertainty in the exposure required to reach a given experimental sensitivity, and undermines any attempt to extract the neutrino mass scale from a measured half-life. 
    Reducing this spread is critical for guiding the design and target selection of future experiments.  More fundamentally, it is essential for discriminating light-neutrino exchange, the standard mechanism mediated by light Majorana neutrinos, from competing lepton-number-violating operators~\cite{Hyvarinen:2015bda}, whose relative contributions can only be disentangled if NMEs are known reliably across several isotopes.

    The nuclear shell model (NSM) is one of the principal many-body methods used to compute such nuclear matrix elements across the nuclear chart~\cite{Caurier2005, shimizu2022recent}.
    Such computations are enabled by efficient NSM codes which are based on iterative Lanczos diagonalization \cite{Sternberg_2008, Maris_2010, Aktulga_2011, johnson2013factorization, johnson2018bigstick, shimizu2013nuclear, shimizu2019thick},
    and have been used to study nuclear structure \cite{cirigliano2022next, brokemeier2025quantum}. 
    Another class of computational methods is the selected configuration interaction (SCI) \cite{roth2009importance, holmes2016heat, li2018fast}.
    Based on finding important configurations of many-body states, SCI algorithms have been used extensively for NSM calculations \cite{stumpf2016importance, wang2026semistochastic, li2026adaptive}.
    Yet the size of the many-body Hilbert space grows exponentially with the number of nucleons considered, motivating the exploration of quantum computing techniques to tackle this scaling problem.
    
    This effort has been pursued extensively.
    With several exceptions known to us \cite{stetcu2023projection, rule2024simplified, costa2025quantum, costa2025quasiparticle, yang2025shadow, yoshida2026nuclear, benstead2026fault}, most of the literature is dedicated to the usage of the variational quantum eigensolver (VQE) algorithm \cite{dumitrescu2018cloud, di2021improving, romero2022solving, perez2023nuclear, bhoy2024shell, gibbs2025exploiting, carrasco2026comparison, kiss2022quantum, stetcu2022variational, sarma2023prediction, yoshida2024accurate, hobday2025variance, singh2025advancing, yoshida2026bridging, rethinasamy2026neutron}.
    Despite successful small-scale demonstrations, VQE has known issues such as flat optimization landscapes \cite{mcclean2018barren, larocca2025barren} and measurement cost \cite{gonthier2022measurements} that present challenges to scaling it beyond what can be achieved classically.  

    A non-variational alternative is sample-based quantum diagonalization (SQD) \cite{robledo2025chemistry}.
    Inspired by classical SCI algorithms,
    the SQD uses a quantum computer to sample basis vectors which form a compact subspace where a target Hamiltonian can be diagonalized.
    Namely, for, e.g., fermionic lattice or molecular Hamiltonians, one samples Slater determinants by preparing and measuring a quantum state which is supposed to yield important configurations.
    Such sampled determinants may be erroneous (e.g., indicate a wrong particle number) due to the noise of the quantum machine.
    To mitigate these errors, the authors of~\cite{robledo2025chemistry} propose the configuration recovery (CR) algorithm, a technique for repairing incorrect sampled determinants using the correct ones. 
    Recent work using synthetic Hamiltonians has shown that SQD can succeed in regimes where widely used classical SCI methods fail~\cite{kirby2026observation} motivating the search for problems for which sampled-based quantum eigensolvers may outperform classical
    approaches.

    In this work, we apply SQD to nuclear Hamiltonians $\Hnuc$ within NSM on the \texttt{ibm\_basquecountry} quantum computer.
    We follow the methodology of the SqDRIFT algorithm proposed in \cite{piccinelli2025quantum}, which combines SQD \cite{robledo2025chemistry} itself, Krylov subspace diagonalization \cite{motta2024subspace, yoshioka2025krylov, yu2025quantum}, and stochastic unitary compilation \cite{campbell2019random}.
    Concretely, Slater determinants are sampled by measuring the Krylov subspace vectors prepared by evolving an initial state under $\Hnuc$, with the evolution approximated using the qDRIFT algorithm \cite{campbell2019random}.
    To compose the evolution operators, we use efficient quantum circuits suitable for simulating fermionic systems \cite{yordanov2020efficient} such as atomic nuclei.
    Additionally, we propose a CR algorithm which repairs noisy sampled Slater determinants, taking into account the symmetries of nuclear Hamiltonians $\Hnuc$.

    We use SQD for approximating the ground state energies of the isotopes $^{48}$Ca, $^{28}$Si, and $^{60}$Zn described by Hamiltonians of 20, 24, and 44 qubits, respectively.
    We first gather both the correct sampled Slater determinants and the noisy determinants repaired with the CR algorithm, and then extend the resultant subspace by classical post-processing \cite{barison2025quantum}.
    This allows to achieve the ground state energies equal to the ones obtained by exact diagonalization.

    The paper is structured as follows.
    Section~\ref{sec:nuclear_shell_model} introduces the NSM and nuclear Hamiltonians $\Hnuc$.
    Section~\ref{sec:methods} describes SQD and other methods used for eigensolving $\Hnuc$ with a quantum computer.
    The results obtained by means of these methods are reported in Section~\ref{sec:results} and discussed in more detail in Section~\ref{sec:discussion}.
    Section~\ref{sec:conclusion} concludes the paper.


\section{Nuclear shell model }
\label{sec:nuclear_shell_model}

    Within the framework of the NSM, also known as the configuration interaction method, the microscopic treatment of the nucleus demands finding an effective nuclear Hamiltonian $H_{\mathrm{nuc}}$.
    It is defined in a reduced space ($\mathcal{H}_{\mathrm{model}}$) that reproduces a subset of the exact eigenstates of a Hamiltonian $H$ in the full, typically intractable Hilbert space ($\mathcal{H}$):
    \begin{equation}
        \label{eq:full_to_model}
        H\ket{\Psi_n} = E_n\ket{\Psi_n}
        \quad\longrightarrow\quad
        H_ {\mathrm{nuc}}\ket{\psi_n} = E_n\ket{\psi_n}.
    \end{equation}
    In this formalism, the single-particle basis is partitioned into three distinct categories: core, valence, and excluded orbits. The model space $\mathcal{H}_{\mathrm{model}}$ is then spanned by Slater determinants in which all core orbitals remain fully occupied (forming an inert core), all excluded orbitals remain empty, and the dynamics of the active nucleons are explicitly treated within the valence space.

    An essential condition for the feasibility of large-scale shell-model calculations is that the effective interaction exhibits a rapidly converging cluster hierarchy.
    That is, the overall strength of the interaction decreases sharply as the number of interacting nucleons increases.
    This way one can safely truncate the system at the two-body level while maintaining high precision and computational tractability. 
    In the configuration space, the nuclear Hamiltonian can then be expressed as a sum of one- plus two-body terms
    \begin{equation}
        \label{eq:H_nuc}
        H_{\mathrm{nuc}} = \sum_\alpha \epsilon_\alpha a^\dagger_\alpha a_\alpha + \frac{1}{4}\sum_{\alpha\beta\gamma\delta} \overline{v}_{\alpha \beta \gamma \delta} a^\dagger_\alpha a^\dagger_\beta a_\delta a_\gamma,
    \end{equation}
    where $\epsilon_\alpha$ is the single-particle energy of the state $\alpha$, and $\overline{v}_{\alpha \beta \gamma \delta}=v_{\alpha \beta \gamma \delta}-v_{\alpha \beta \delta \gamma}$ are the anti-symmetrized and normalized two-body matrix elements (TBMEs). 
    The fermionic operator $a^\dagger_\alpha$ ($a_\alpha$) creates (annihilates) a nucleon in the single-particle orbital $\alpha$,
    \begin{equation}
        \label{eq:M-scheme_nucleon_states}
        \ket{\alpha} \equiv |n^{(\alpha)}, l^{(\alpha)}, j^{(\alpha)}, m^{(\alpha)}, t_z^{(\alpha)}\rangle \, ,
    \end{equation}
    where $n$, $l$, and $j = l \pm \frac12$ denote the principal, orbital, and total angular momentum quantum numbers, $m \in \{-j, \ldots, j\}$ is the projection of $j$, and the isospin projection $t_z = \pm\frac12$ distinguishes protons ($-\frac12$) from neutrons ($+\frac12$).
    The energy levels of different orbitals\footnote{From here on, if not stated otherwise, when we refer to ``orbitals'' we mean single-particle orbitals \eqref{eq:M-scheme_nucleon_states}.}
    align into groups, or shells, typically separated by large energy gaps, hence the name ``nuclear shell model''.
    Nuclear wavefunctions are written as superpositions of Slater determinants, correctly antisymmetrized products of the single-nucleon eigenstates~\eqref{eq:M-scheme_nucleon_states}.
    
    The matrix elements $\overline{v}_{\alpha \beta \gamma \delta}$ can be derived from an effective field theory rooted in quantum chromodynamics~\cite{Stroberg19}, the underlying theory of the nuclear force. 
    In this work, however, we employ phenomenological Hamiltonians whose parameters are adjusted to reproduce available experimental data, such as the binding energies and excitation spectra of selected nuclei.
    The TBMEs for such Hamiltonians are obtained for the orbitals of certain shells associated with sets of quantum numbers $nl_j$ constituting the model space.

    As an illustrative example, Table~\ref{tab:o2q-ckpot} shows the model space for the Cohen-Kurath interaction \cite{cohen1965effective}. 
    This interaction describes nucleons in the $p$-shell with $n=0$, meaning that the orbital angular momentum is $l=1$ (or $p$ in spectroscopic notation) giving $j \in \{\frac12, \frac32\}$ and thus the orbitals $0p_{1/2}$ and $0p_{3/2}$. 
    Along with the remaining quantum numbers $m \in \{-j, \ldots, j\}$ and the two nucleon species distinguished by the isospin projection $t_z = \pm \frac12$, this results in a model space of $\norb = 12$ single-particle states \eqref{eq:M-scheme_nucleon_states}. 
    As the states in the $0s_{1/2}$ orbital (two for protons and two for neutrons) are assumed to be occupied, the Cohen-Kurath interaction therefore effectively describes the nucleons in the $p$-shell above the $^{4}$He core.

    \begin{table}
        \centering
        \caption{
            Model space 
            for the $p$-shell. 
        }
        \begin{tabular}{|c|cc|cccc||cc|cccc|}
            \hline
            $\alpha$ & $1$  & $2$  & $3$  & $4$  & $5$  & $6$  & $7$  & $8$  & $9$  & $10$  & $11$ & $12$ \\
            \hline
            $n$      & $0$  & $0$  & $0$  & $0$  & $0$  & $0$  & $0$  & $0$  & $0$  & $0$  & $0$   & $0$ \\
            $l$      & $p$  & $p$  & $p$  & $p$  & $p$  & $p$  & $p$  & $p$  & $p$  & $p$  & $p$   & $p$ \\
            $2j$     & $1$  & $1$  & $3$  & $3$  & $3$  & $3$  & $1$  & $1$  & $3$  & $3$  & $3$   & $3$ \\
            $2m$     & $-1$ & $+1$ & $-3$ & $-1$ & $+1$ & $+3$ & $-1$ & $+1$ & $-3$ & $-1$ & $+1$  & $+3$ \\
            $2t_z$   & $+1$ & $+1$ & $+1$ & $+1$ & $+1$ & $+1$ & $-1$ & $-1$ & $-1$ & $-1$ & $-1$  & $-1$ \\
            \hline
        \end{tabular}
        \label{tab:o2q-ckpot}
    \end{table}

    As the single-nucleon states \eqref{eq:M-scheme_nucleon_states} have distinct single-nucleon total angular momentum projections $m^{(\alpha)}$, they couple to states with well-defined two-nucleon total angular momentum projections $M = m^{(\alpha)} + m^{(\beta)}$.
    The TBMEs $\overline{v}_{\alpha \beta \gamma \delta}$ in \eqref{eq:H_nuc} with such a coupling are said to be represented in the $M$-scheme, which is the representation used throughout this work.
    As will be seen in the next section, this representation is convenient for quantum computations. 
    However, the TBMEs are commonly given in the more compact $J$-scheme, where the single-nucleon states are coupled to have well-defined total angular momentum $|j^{(\alpha)} - j^{(\beta)}| \leqslant J \leqslant j^{(\alpha)} + j^{(\beta)}$.
    In Appendix~\ref{app:J-to-M}, we describe the transformation of TBMEs from the $J$- to the $M$-scheme.

    In this work, we aim to use a quantum computer to (approximately) solve the ground state problem
    \begin{equation*}
        (E_{\mathrm{gs}},\ket{\psi_{\mathrm{gs}}}) = \arg\min_{E,\ket{\psi}} \left\{ E : H_{\mathrm{nuc}}\ket{\psi} = E\ket{\psi} \right\},
    \end{equation*}
    i.e., we want to find the lowest eigenvalue $E_{\mathrm{gs}}$ and the corresponding eigenstate $\ket{\psi_{\mathrm{gs}}}$ of $H_{\rm nuc}$.

    
\section{Methods}
\label{sec:methods}

    In this section, we describe the methods we use for obtaining an approximate ground state energy $\tilde{E}_{\mathrm{gs}}$ of a nuclear Hamiltonian $\Hnuc$ of the form \eqref{eq:H_nuc} on a quantum computer.
    We do this by using the SqDRIFT algorithm \cite{piccinelli2025quantum} and perform the following steps. 
    \begin{enumerate}
        \item Define an orbital-to-qubit mapping for $\Hnuc$ (see Section~\ref{sec:orbital-to-qubit}).
        \item Pick an initial state and prepare Krylov states using the qDRIFT algorithm (see Sections~\ref{sec:skqd}, \ref{sec:initial_state}, \ref{sec:qdrift}).
        \item Sample bit strings $\mathcal{S}$ by measuring the Krylov states in the computational basis, determine the correct $\mathcal{C} \subset \mathcal{S}$ and incorrect $\mathcal{I} \subset \mathcal{S}$ bit strings, based on whether they satisfy the required symmetries.
        \item Execute the configuration recovery algorithm to repair $\mathcal{I}$ and obtain the recovered bit strings $\mathcal{R}$ (see Section~\ref{sec:cr}).
        \item Apply classical post-processing to $\mathcal{C} \cup \mathcal{R}$ to get an extended set of correct bit strings $\mathcal{E}$ (see Section~\ref{sec:extension}).
        \item Project $\Hnuc$ onto $\mathcal{E}$ and eigensolve it in this subspace to obtain $\tilde{E}_{\mathrm{gs}}$ (see Equation~\ref{eq:H_proj-eigenproblem}).
    \end{enumerate}
    In the following subsections, we cover each of these steps in detail.


    \subsection{Mapping nucleon orbitals to qubits}
    \label{sec:orbital-to-qubit}

        Given a nuclear Hamiltonian $\Hnuc$, each of the $\norb$ single-particle orbitals specified by the quantum numbers
        \begin{equation}
            \label{eq:orbital_quantum_numbers}
            \alpha = (n^{(\alpha)}, l^{(\alpha)}, j^{(\alpha)}, m^{(\alpha)}, t_z^{(\alpha)})
        \end{equation}
        is assigned to a qubit via the Jordan-Wigner mapping.
        Slater determinants defining a basis for many-nucleon states are written as computational basis $\{\ket{s}\}_{s \in \{0, 1\}^{\times n_\mathrm{q}}} = \{\ket{0}, \ket{1}\}^{\otimes n_\mathrm{q}}$ denoted by bit strings $s = s_1 s_2 \ldots s_{n_\mathrm{q}}$.
        The value of each bit $s_\alpha \in \{0, 1\}$ indicates the occupation of the orbital $\alpha \in \{1, 2, \ldots, n_\mathrm{q}\}$ indexing the five quantum numbers as in \eqref{eq:orbital_quantum_numbers}.

        As an example, let us consider the $^{6}$Li isotope.
        Within the Cohen-Kurath interaction model space introduced in Section~\ref{sec:nuclear_shell_model}, it has $\npro = 1$ valence proton and $\nneu = 1$ valence neutron above the $^{4}$He core.
        The model space shown in Table~\ref{tab:o2q-ckpot} can also serve as an orbital-to-qubit mapping.
        With this mapping, one needs $n_\mathrm{q}=12$ qubits to describe a many-nucleon state.
        One such state for $^{6}$Li could be represented by a bit string $s = 10 \svert 0000 \dsvert 01 \svert 0000$, where we use $\dsvert$ to separate protons and neutrons, and $\svert$ to distinguish orbitals with the same $j$.
        This bit string indicates a neutron being in the state $\ket{0p_{1/2}, m=-1/2, t_z=1/2}$, and a proton in $\ket{0p_{1/2}, m=1/2, t_z=-1/2}$.  

        Therefore, the ground state $\ket{\psi_\mathrm{gs}}$ of \eqref{eq:H_nuc} can now be represented as a linear combination of such bit strings:
        \begin{equation}
            \label{eq:gs_bitstrings}
            \ket{\psi_\mathrm{gs}} = \sum_{s\in \{0,1\}^{\times n_q}} g_s^{\mathrm{(gs)}} \ket{s}.
        \end{equation}
        In addition to the quantum numbers \eqref{eq:orbital_quantum_numbers} associated with the bits $s_\alpha$, each bit string $s$ can be characterized by the following quantities.
        \begin{itemize}
            \item Particle number $n_{\mathrm{nuc}} = \sum_{\alpha = 1}^{\norb} s_\alpha$, the Hamming weight of $s$.
            \item Total isospin projection $T_z = \sum_{\alpha = 1}^{\norb} s_\alpha t_z^{(\alpha)}$.
            \item Total angular momentum projection $M = \sum_{\alpha = 1}^{\norb} s_\alpha m^{(\alpha)}$.
            \item Parity $\Pi = (-1)^{\sum_{\alpha = 1}^{\norb} s_\alpha l_\alpha}$.
        \end{itemize}
        Since they commute with $H_\mathrm{nuc}$, the quantities $n_{\mathrm{nuc}}$, $T_z$, $M$ and $\Pi$ are conserved quantities with the shared eigenbasis.
        While not generally true for individual bit strings $\ket{s}$, the whole nuclear ground state $\ket{\psi_\mathrm{gs}}$ also has a well-defined total angular momentum $J$, often indicated together with the parity $\Pi$ as $J^{\Pi}$.
        The number of many-body states with specific $\nnuc$, $T_z$, $M$ and $\Pi$ in a chosen model space is called the $M$-scheme dimension.

        As an example,  the $^{8}$Be isotope within the Cohen-Kurath interaction model space (see Table~\ref{tab:o2q-ckpot}), has $\npro=\nneu=2$ valence particles, which implies $T_z = 0$.
        The ground state is known to have $J^\Pi = 0^+$ giving also $M=0$~\cite{nudat3}.
        Therefore, the ground state of $^{8}$Be can be represented as a combination of bit strings of the Hamming weight $\nnuc = \npro + \nneu = 4$ with the above quantum numbers.
        With the above-mentioned quantum numbers, the $M$-scheme dimension is 51.

        
    \subsection{Sample-based Krylov diagonalization}
    \label{sec:skqd}
    
        Representing states as a linear combination of determinant bit strings, $\ket{\tilde{\psi}}=\sum g_s \ket{s}$, allows one to write approximations of the eigenstates of $\Hnuc$ using a subset of determinants $\mathcal{S} \equiv \{s\}_s$.
        Finding the approximation in the subspace spanned by $\mathcal{S}$ reduces to solving the eigenvalue equation 
        \begin{equation}
            \label{eq:H_proj-eigenproblem}
            \tilde{H}_{\mathrm{nuc}} |\tilde{\psi}\rangle = \tilde{E}|\tilde{\psi}\rangle, \quad 
            \tilde{H}_{\mathrm{nuc}} = \sum_{s,r \in \mathcal{S}} \langle s | \Hnuc | r\rangle \ketbra{s}{r}.
        \end{equation}
        For quartic fermionic Hamiltonians, matrix elements $\langle s | \Hnuc | r\rangle$ for the projected Hamiltonian $\tilde{H}_{\mathrm{nuc}}$ can be computed straightforwardly.
        Indeed, since the fermionic operators $a^\dagger_\gamma$ and $a_\gamma$ act on the bits $s_\gamma \in \{0, 1\}$ of $\ket{s}$ as
        \begin{align}
            \label{eq:cre_des_action}
            \begin{split}
                a^\dagger_\gamma \ket{s} &= (-1)^{\sum_{\alpha=1}^{\gamma - 1} s_\alpha} \, (1 - s_\gamma) \;\ket{\ldots (1 - s_\gamma) \ldots}, \\
                a_\gamma \ket{s} &= (-1)^{\sum_{\alpha=1}^{\gamma - 1} s_\alpha} \, s_\gamma \;\ket{\ldots (1 - s_\gamma) \ldots}, 
            \end{split}
        \end{align}
        one can sequentially apply the terms of $\Hnuc$ from \eqref{eq:H_nuc} to each bit string $s \in \mathcal{S}$ and, knowing that $\langle s'|s\rangle = \delta_{s's}$, obtain the desired matrix elements.

        If $\Hnuc$ has large support on a small $\mathcal{S}$, compared with the $M$-scheme dimension, then such an approximation can closely match the true ground state. 
        This requires that (a) $\Hnuc$ and the determinant basis are such that the true ground state has support concentrated on a small set of determinants, i.e., $|g_s|$ is large for $s\in \mathcal{S}$ and small otherwise, and (b) there exists an efficient algorithm for generating $\mathcal{S}$. 
        While the former is a fundamental property of the physical system and ultimately limits the accuracy given a fixed $|\mathcal{S}|$, solutions for the latter have been studied extensively, particularly for molecular electronic structure problems, using classical computations in SCI approaches~\cite{holmes2016heat, dash2018perturbatively,levine2021cas,zhang2025random} and quantum computation, in quantum-SCI or SQD algorithms~\cite{motta2024subspace, robledo2025chemistry, kanno2026quantum, sugisaki2025hamiltonian, mikkelsen2025quantum}.

        Sample-based quantum Krylov diagonalization (SKQD)~\cite{yu2025quantum} is a particularly powerful variant of SQD. In SKQD, given a reference state $\ket{\psi_0}$, one uses a quantum computer to perform unitary evolution under $\Hnuc$ for a discrete set of times $td$ with $d \in \{0, 1, \ldots, D-1\}$ and $t \in \R$, to prepare a set of states
        \begin{equation}
            \label{eq:krylov_states}
            \mathcal{K}_D \equiv \big\{ e^{-i d \Hnuc t } \ket{\psi_0} \big\}_{d=0}^{D-1}
        \end{equation}
        known as Krylov states \cite{piccinelli2025quantum, yu2025quantum}.
        Measuring the states $\mathcal{K}_D$ in the computational basis produces the bit strings $\mathcal{S}$ onto which one can project and eigensolve the Hamiltonian in question using a classical computer. 
        In the absence of hardware noise, SKQD comes with powerful convergence guarantees dependent on $|\langle \psi_0 | \psigs\rangle|^2$, provided that $\Hnuc$ satisfies the aforementioned property of concentrated support (quantified by a sparsity criterion)~\cite{yu2025quantum}.


    \subsection{Initial state}
    \label{sec:initial_state}

        As the initial state $\ket{\psi_0}$ for preparing the Krylov vectors \eqref{eq:krylov_states},
        we take a lowest-energy Slater determinant as proposed in \cite{costa2025quantum}:
        \begin{equation}
            \label{eq:psi0-finding}
            \psi_0 = \arg\min_{s} \sum_{\alpha=1}^{\norb} s_\alpha (\nu\epsilon_\alpha - |m_\alpha|),
        \end{equation}
        where 
        $\nu > 0$.
        That is, $\ket{\psi_0}$ is represented by a bit string which indicates the nucleons occupying the orbitals $\alpha$ with the lowest single-body energies $\epsilon_\alpha$ and largest (in magnitude) total angular momentum projections $m_\alpha$.
        The factor $\nu$ must be properly chosen to ensure that we first minimize the total single-body energy, and only then maximize the sum of $|m_\alpha|$.
        We also require the state $\ket{\psi_0}$ to have the desired quantum numbers $\nnuc$, $T_z$, $M$, and $\Pi$ introduced in Section~\ref{sec:orbital-to-qubit}.
        This gives rise to the corresponding constraints for the minimization problem \eqref{eq:psi0-finding}.
        For now, we omit these constraints from the equation and discuss them later in Section~\ref{sec:cr}.

        
    \subsection{Stochastic time evolution}
    \label{sec:qdrift}

        We prepare Krylov states~\eqref{eq:krylov_states} following an SqDRIFT~\cite{piccinelli2025quantum} approach, a variant of SKQD that approximates the unitary time evolution via a stochastic evolution called qDRIFT~\cite{campbell2019random}. 
        Using qDRIFT, evolution under the Hamiltonian $\Hnuc$ is approximated as
        \begin{equation}
            \label{eq:qdrift}
            e^{- i H_{\mathrm{nuc}}t} \approx \prod_{\kappa=1}^K \exp\left(-i \, \mathrm{sign}(e_{l_\kappa}) \frac{t \lambda}{K} E_{l_\kappa} \right),
        \end{equation}
        where $E_{l}$ are one- and two-body operators taken from \eqref{eq:H_nuc}, i.e.,\footnote{Here the orbitals $\alpha$ are indexed by $l$, i.e., formally, $\alpha \equiv \alpha_l$. We omit this dependence for readability.}
        \begin{equation}
            \label{eq:E_operators}
            E_{l} = a^\dagger_\alpha a_\alpha \quad \text{ or } \quad E_{l} = a^\dagger_\alpha a^\dagger_\beta a_\delta a_\gamma + \mathrm{H.c.},
        \end{equation}
        where we exclude the Hermitian conjugate from the latter if $\alpha = \gamma, \beta = \delta$ or $\alpha = \delta, \beta = \gamma$ to avoid double counting.
        Each of the $K$ operators $E_{l_k}$ is chosen randomly with the probability $p_l = |e_l|/\lambda$, where $e_l = \epsilon_\alpha$  or $e_l = \overline{v}_{\alpha \beta \gamma \delta}/4$, and $\lambda = \sum_{l} |e_l|$.
        In our setting, simulating the Hamiltonian as in \eqref{eq:qdrift} ensures that the quantum numbers $\nnuc$, $T_z$, $M$ and $\Pi$ of $\ket{\psi_0}$ are conserved.\footnote{The TBMEs $\overline{v}_{\alpha \beta \gamma \delta}$ in the Hamiltonian \eqref{eq:H_nuc} connect only two-nucleon states $\ket{\alpha \beta}$ and $\ket{\gamma \delta}$ with the same $T_z$, $M$, and $\Pi$.}
        The simulation error of qDRIFT scales as $O(\lambda^2 t^2/K)$ \cite{campbell2019random, chen2020concentration}.

        A simple approach to simulating the evolution of a Hamiltonian---as well as of the individual terms $E_l$ in \eqref{eq:E_operators}---would be representing it in the Pauli basis via the Jordan-Wigner transformation, and then simulating each Pauli string separately. 
        However, one may recognize the exponentials of $E_{l}$ in \eqref{eq:qdrift} to be fermionic excitation operators, which admit efficient quantum circuit representations \cite{yordanov2020efficient}.
        Such circuits have lower counts of the two-qubit gates compared to simulating individual Pauli strings.
        In Appendix~\ref{app:fermionic_gates}, we describe and discuss in detail the circuits for $\exp(-i \theta E_l)$, $\theta \in \mathbb{R}$,  we use in our work.

        Additionally, to improve the Hamiltonian simulation, we omit from the circuit the operators that act trivially at the beginning of the circuit and use the symmetry shift method proposed in Ref.~\cite{loaiza2023reducing}. 
        Further details are given in Appendix~\ref{app:qdrift_improvements}.


\subsection{Configuration recovery}
    \label{sec:cr}

        Although the circuits we use for preparing the (approximate) Krylov vectors $e^{-i d \Hnuc t } \ket{\psi_0}$ preserve the quantum numbers $\nnuc$, $T_z$, $M$, and $\Pi$ contained in $\ket{\psi_0}$, the noise of the quantum computer may corrupt the Krylov states and bit strings $s \in \mathcal{S}$ sampled by measuring them.
        These noisy, incorrect bit strings $\mathcal{I} \subset \mathcal{S}$ with wrong quantum numbers can simply be discarded when projecting and eigensolving the Hamiltonian as in \eqref{eq:H_proj-eigenproblem}.

        To avoid such a waste of quantum resources, SQD~\cite{robledo2025chemistry} proposes the configuration recovery (CR) algorithm, a technique that uses the correct bit strings $\mathcal{C} = \mathcal{S} \setminus \mathcal{I}$ for fixing the incorrect ones $\mathcal{I}$.
        The original CR algorithm was developed for electronic quantum chemistry Hamiltonians, and is based on probabilistically flipping the bits of incorrect strings to restore the numbers of spin-up and spin-down electrons.

        As we discuss in Appendix~\ref{app:cr_algorithm}, this algorithm is not directly applicable to nuclear shell model Hamiltonians \eqref{eq:H_nuc}.
        This is because quantum numbers $M$ and $\Pi$ put additional restrictions on which bits can be flipped in an incorrect bit string $i \in \mathcal{I}$ to restore the quantum numbers $\nnuc$, $T_z$, $M$, and $\Pi$. 
        While these restrictions make implementation of a CR-like algorithm more conceptually challenging, the additional symmetries may lead to increased probability of recovering a bit string corresponding to large $g_s^{\mathrm{(gs)}}$ from highly corrupted ones.
        To find the bits that can be flipped for repairing a bit string, one needs to solve (a version of) the subset sum problem, which admits a linear programming implementation similar to \eqref{eq:psi0-finding}.  

        Concretely, given an incorrect bit string $i \in \mathcal{I}$ with bits $i_\alpha \in \{0, 1\}$, we find a recovered bit string $r(i)$ by solving
        \begin{align}
            \label{eq:recovered_bitstring}
            \begin{split}
                &r(i) = \arg\max_{s} \sum_{\alpha=1}^{\norb} s_\alpha  i_{\alpha} \\
                &\text{s.t. } \sum_{\alpha=1}^{\norb} s_\alpha = \nnuc, \qquad\quad\; \sum_{\alpha=1}^{\norb} s_\alpha t_z^{(\alpha)} = T_z, \\
                &\phantom{s.t.\;\,} \sum_{\alpha=1}^{\norb} s_\alpha m^{(\alpha)} = M, \qquad (-1)^{\sum_{\alpha=1}^{\norb} s_\alpha l^{(\alpha)}} = \Pi.
            \end{split}
        \end{align}
        That is, by varying over bit strings $s$ with bits $s_\alpha \in \{0, 1\}$, we maximize the overlap of $s$ with the given $i$.
        This maximization is subject to the constraints ensuring that the recovered bit string $r(i)$ has the correct quantum numbers $\nnuc$, $T_z$, $M$, and $\Pi$.
        In other words, $r(i)$ is found as a correct bit string closest to the given incorrect one $i$ by Hamming distance.
        We call this approach of repairing bit strings for nuclear Hamiltonians the nuc-CR algorithm.
        
        In \eqref{eq:recovered_bitstring}, the only equation that is not linear in the binary variables $s_\alpha \in \{0, 1\}$ is the constraint on the parity $\Pi$.
        In Appendix~\ref{app:cr_algorithm}, we show that it can be linearized by introducing an auxiliary integer variable.
        This reduces \eqref{eq:recovered_bitstring} to a linear programming problem solvable using packages such as PuLP \cite{pulp} employed in this work.


    \subsection{Extending the set of correct bit strings} 
    \label{sec:extension}
    
        After repairing the incorrect bit strings, one can use classical post-processing to extend the set of the correct ones $\mathcal{B} = \mathcal{C} \cup \mathcal{R}$ by performing certain bit-wise operations.
        First, in the case when $M=0$, we can ``time-reverse'' \cite{costa2025quasiparticle, yoshida2026bridging} the bit strings by swapping the bits such that $m \rightarrow -m$.
        That is, we reverse the order of bits within the sectors of distinct values of $j$.
        For instance, consider again the isotope $^{8}$Be mentioned in Section~\ref{sec:orbital-to-qubit}.
        A possible state of this isotope can be represented by a bit string $s = 00 \svert 0011 \dsvert 10 \svert 1000$.
        After the time-reversal, this bit string becomes $s' = 00 \svert 1100 \dsvert 01 \svert 0001$, which has the same quantum numbers $\nnuc$, $T_z$, $M$, and $\Pi$, i.e., is a valid bit string that can be used for the projection and diagonalization in \eqref{eq:H_proj-eigenproblem}.

        Second, taking inspiration from classical SCI algorithms \cite{holmes2016heat, li2018fast} and following recent works \cite{piccinelli2025quantum, barison2025quantum}, one can extend the set of correct bit strings by applying the excitation terms from the Hamiltonian itself.  
        That is, one takes the terms $a^\dagger_\alpha a^\dagger_\beta a_\delta a_\gamma$ from \eqref{eq:H_nuc} and applies them to a (correct) sampled Slater determinant $\ket{s}$, which effectively results in swapping the digits of the bit string $s$.
        As an example, consider $\ket{s} = \ket{00 \svert 0011 \dsvert 10 \svert 1000}$ represented by the bit string $s$ mentioned earlier, and the operator $a^\dagger_2 a^\dagger_{10} a_7 a_5$ taken from \eqref{eq:H_nuc}.
        One can therefore obtain a new correct bit string $s''$ as $\ket{s''}=a^\dagger_2 a^\dagger_{10} a_7 a_5 \ket{s} \simeq \ket{01 \svert 0001 \dsvert 00 \svert 1100}$.\footnote{The equality here is up to a sign induced by the anticommutation relations of $a^\dagger$ and $a$ as in \eqref{eq:cre_des_action}, which is not relevant here for obtaining valid bit strings.}

        The two techniques described above can be applied to $\mathcal{B}$ several times iteratively to get an extended set of correct bit strings $\mathcal{E}$.
        After each application, one can assess the importance of the new bit strings and discard the less important ones between iterations, as commonly done in SCI algorithms.


\section{Results}
\label{sec:results}

    In this section, we report the results of the SqDRIFT algorithm executed on the \texttt{ibm\_basquecountry} quantum computer.
    We applied the algorithm to find the ground state energy for the isotopes $^{48}$Ca, $^{28}$Si, and $^{60}$Zn.
    The model spaces used for these isotopes, as well as the numbers of valence particles $\npro$ and $\nneu$ and the quantum numbers $T_z$, $M$, and $J^\Pi$ of the ground state are outlined in Table~\ref{tab:isotopes}.  

    The isotope $^{48}$Ca, requiring $n_\mathrm{q}=20$ qubits, is a particularly interesting benchmark for neutrinoless double-beta decay. It is the target of the CANDLES experiment~\cite{Umehara2008}, and was previously studied by CARVEL~\cite{Zdesenko2005}, while also being the subject of significant theoretical calculations of its nuclear matrix elements~\cite{Iwata2016,Horoi2010,Yao2020,Belley2021,Novario2021}. 
    The ground state of the $^{28}$Si isotope consists of a (relatively) large number of bit strings of length $n_\mathrm{q}=24$, making it a suitable test case for the algorithm.
    The model space used for the isotope $^{60}$Zn demands $n_\mathrm{q} = 44$ qubits, requiring hardware for larger-scale quantum computations.

    \begin{table*}
        \centering
        \caption{
            Isotopes studied in this work, the model spaces used, number of qubits $n_\mathrm{q}$ needed, numbers of valence protons $\npro$ and neutrons $\nneu$, quantum numbers $T_z$, $M$, and $J^\Pi$ of the ground state, and the $M$-scheme dimensions.
            The last two columns show the two-qubit gate counts and depths of the qDRIFT circuits presented as mean $\pm$ standard deviation.
        }
        \begin{tabular}{|c|c|c|c|c|c|c|c|c|c|c|}
            \hline
            Isotope    & Shell (interaction)              & $n_\mathrm{q}$ & $\npro$ & $\nneu$ & $T_z$ & $J^\Pi$ & $M$ & $M$-scheme dimension & 2-qubit gate depth  & 2-qubit gate count  \\
            \hline
            $^{48}$Ca  & $pf$ (KB3G \cite{poves2001shell}) & 20             & 0       & 8       & $+4$  & $0^+$   & 0   & 12022          & $347.31 \pm 76.98$  & $ 521.13 \pm 116.03$ \\
            $^{28}$Si  & $sd$ (USDB \cite{brown2006new})   & 24             & 6       & 6       & $0$   & $0^+$   & 0   & 93710          & $303.68 \pm 84.62$  & $481.17 \pm  131.41$ \\
            $^{60}$Zn & $fpg$ (JUN45 \cite{honma2009new})  & 44             & 2       & 2       & $0$   & $0^+$   & 0   & 2533           & $489.98 \pm 160.25$ & $795.82 \pm 237.17$ \\
            \hline
        \end{tabular}
        \label{tab:isotopes}
    \end{table*}


    \subsection{SqDRIFT settings}
    
        We executed the SqDRIFT algorithm with the following settings.
        The time step indices in the Krylov states \eqref{eq:krylov_states} are $d \in \{1, \ldots, 5\}$,\footnote{Note that we deliberately skipped $d=0$, which results in the Krylov vector being the initial state $\ket{\psi_0}$, a single bit string.} 
        and the evolution time itself was set to $t=1$.
        Each Krylov state was prepared $R=50$ times with different sequences of operations sampled for the qDRIFT circuit \eqref{eq:qdrift}. 
        Each circuit contained $K=10$ randomly sampled operators $e^{-i \theta E_l}$.
        The numbers and depths of the two-qubit gates after transpilation of these circuits for \texttt{ibm\_basquecountry} are shown in Table~\ref{tab:isotopes}.
        Measuring each of the $R=50$ qDRIFT randomizations of the 5 Krylov states 1000 times produced $|\mathcal{S}| = 250,\!000$ (non-unique) bit strings that are processed as described below.


    \subsection{Obtaining the bit strings}
    \label{sec:bitstrings-obtained}

        \begin{table}
            \centering
            \caption{
                Numbers of correct bit strings (Slater determinants) obtained at different acquisition stages.
            }
            \begin{tabular}{|c|c|c|c|c|}
                \cline{2-5}
                \multicolumn{1}{c}{} & \multicolumn{4}{|c|}{Number of correct determinants obtained}  \\
                \hline
                Isotope    & Sampling & Recovery & Time-reversal & Excitation \\
                \hline
                $^{48}$Ca  &     2138 &    11748 &         11992 &      12022 \\
                $^{28}$Si  &     2118 &    47143 &         62778 &      93710 \\
                $^{60}$Zn &       95 &     2533 &          2533  &      2533 \\
                \hline
            \end{tabular}
            \label{tab:three_isotopes-bitstrings}
        \end{table}
        
\begin{figure*}
            \centering
            \subfloat[\label{fig:three_isotopes-bitstrings}]{
                \includegraphics[width=0.477\linewidth]{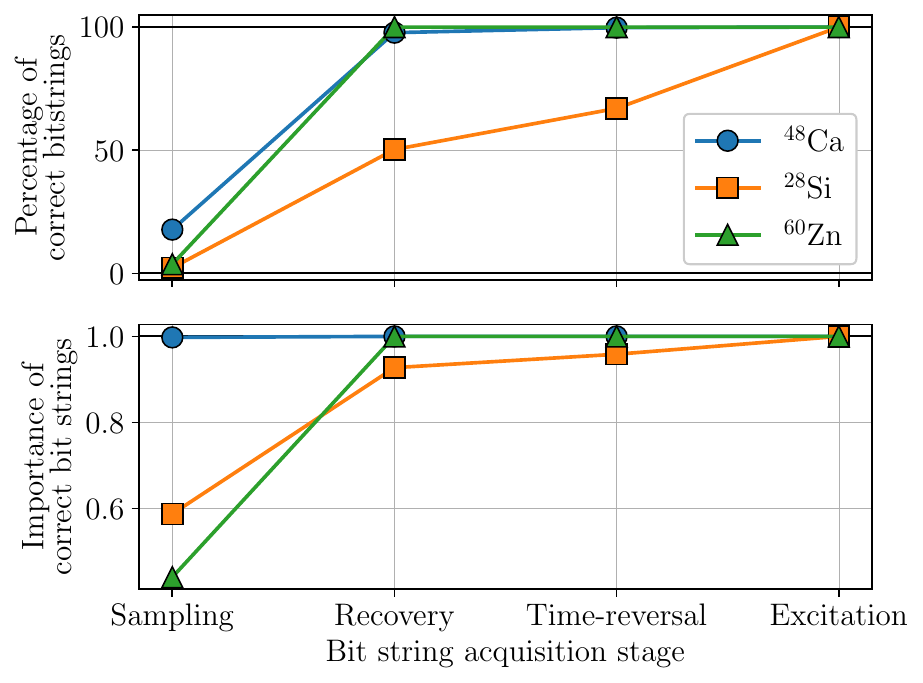}
            }
            \hfill 
            \subfloat[\label{fig:three_isotopes-energy_errors}]{
                \includegraphics[width=0.485\linewidth]{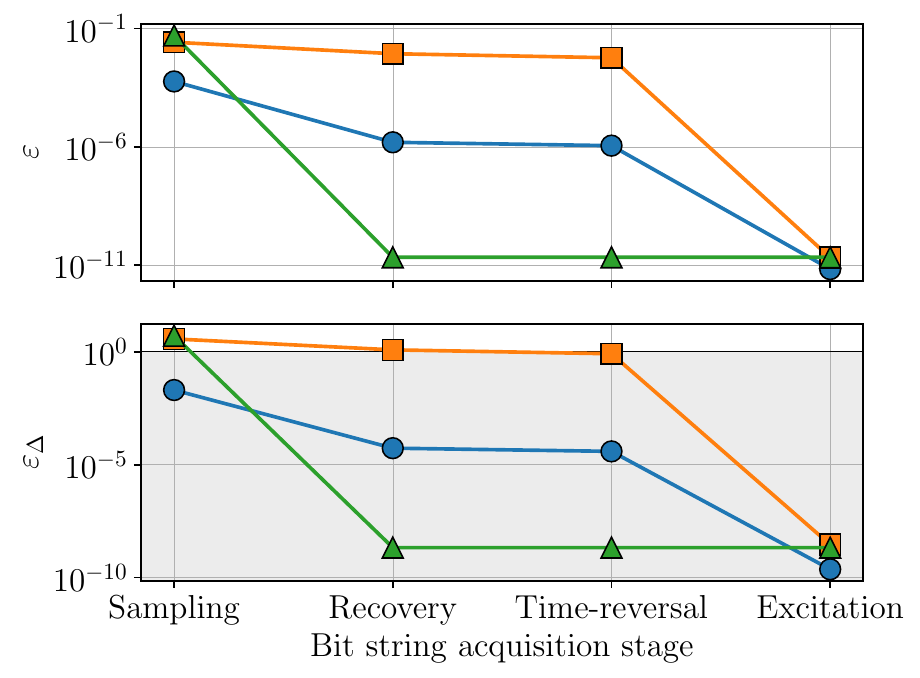}
            }
            \caption{
                Results of (a) obtaining Slater determinants and (b) ground state energy approximation for $^{48}$Ca, $^{28}$Si, $^{60}$Zn.
                The results are shown for the four bit string acquisition stages involving varying amounts of additional classical post-processing: 
                using raw samples from the quantum computer postselected to have the required quantum numbers,
                adding bit strings obtained by executing nuc-CR for the incorrect raw samples (see Section~\ref{sec:cr}),
                and extending the subspace formed in the previous steps by application of the time-reversal and excitations to the bit strings (see Section~\ref{sec:extension}).
                The upper panel of (a) shows the percentage of the correct bit strings obtained, while the lower panel plots their importance assessed as their coefficients in the exact ground state \eqref{eq:gs_bitstrings}.
                The errors of the approximate ground state energy $\tilde{E}_{\mathrm{gs}}$ shown in (b) are expressed via the quantities \eqref{eq:rel_error} and \eqref{eq:gap_error};
                the shaded area in the lower panel indicates $\varepsilon_\Delta < 1$.
            }
            \label{fig:three_isotopes-bitstrings-energy_errors}
        \end{figure*}

        We split the acquisition of correct bit strings for projecting and eigensolving $\Hnuc$ into four stages: (i) sampling from the quantum computer, (ii) executing the configuration recovery algorithm for the incorrect bit strings, and applying (iii) time-reversal and (iv) excitations to the correct and corrected bit strings.
        Table~\ref{tab:three_isotopes-bitstrings} shows the numbers of correct bit strings obtained at each of these stages for the studied isotopes $^{48}$Ca, $^{28}$Si, and $^{60}$Zn.
        In the upper panel of Figure~\ref{fig:three_isotopes-bitstrings}, we also plot these numbers as percentages of the $M$-scheme dimensions as per Table~\ref{tab:isotopes}. 
        The lower panel of this figure shows the importance of the bit strings assessed as $\sum_{s} |g_s^{\mathrm{(gs)}}|^2$, where $g_s^{\mathrm{(gs)}}$ are the coefficients of the corresponding bit strings $s$ in the exact ground state \eqref{eq:gs_bitstrings}.

        By looking at Table~\ref{tab:three_isotopes-bitstrings} and Figure~\ref{fig:three_isotopes-bitstrings}, one can see that we sample relatively few initially correct bit strings $\mathcal{C}$ from the quantum hardware.
        However, these bit strings are important by the metric mentioned above.
        Indeed, for $^{28}$Si and $^{60}$Zn the total importance of the sampled strings is about 0.59 and 0.44, respectively, while for $^{48}$Ca it is more than 0.99.
        As we will see in the next section, this also implies a low error of the ground state energy approximation.

        Executing the nuc-CR algorithm for the incorrect sampled bit strings $\mathcal{I}$ produces more than half of Slater determinants from the exact ground state for $^{28}$Si, almost all the determinants for $^{48}$Ca, and exactly all for $^{60}$Zn.
        Applying the time-reversal and excitations to the correct and corrected bit strings allows to obtain all the bit strings from the ground state.

    \subsection{Projecting and eigensolving $\Hnuc$}

        We can finally project our Hamiltonian onto the subspace spanned by the obtained bit strings and eigensolve it there, as described by \eqref{eq:H_proj-eigenproblem}.
        To assess the quality of the approximate ground state energy $\tilde{E}_{\mathrm{gs}}$ obtained as a solution, we use the following two metrics.
        The first one is the relative error
        \begin{equation}
            \label{eq:rel_error}
            \varepsilon = \frac{\tilde{E}_{\mathrm{gs}} - E_{\mathrm{gs}}}{|E_{\mathrm{gs}}|}
        \end{equation}
        commonly reported in the literature.
        The second metric we use is
        \begin{equation}
            \label{eq:gap_error}
            \varepsilon_\Delta = 2\frac{\tilde{E}_{\mathrm{gs}} - E_{\mathrm{gs}}}{E_{\mathrm{1ex}} - E_{\mathrm{gs}}},
        \end{equation}
        where $E_{\mathrm{1ex}}$ is the true\footnote{By ``true'' or ``exact'' eigenstate and its energy in this work we mean those obtained by exact diagonalization of $\Hnuc$.}
        energy of the first excited state.
        This quantity shows the error in relation to a half of the energy gap $\Delta \equiv E_{\mathrm{1ex}} - E_{\mathrm{gs}}$, so $\varepsilon_\Delta < 1$ means that we are closer to the ground state than to the first excited one. 

        In Figure~\ref{fig:three_isotopes-energy_errors}, we show the energy errors \eqref{eq:rel_error} and \eqref{eq:gap_error} for each of the bit string acquisition stages discussed in the previous section.
        As one can see, for $^{48}$Ca we sample enough bit strings to achieve $\varepsilon_\Delta \approx 2 \cdot 10^{-2}$, signifying that the energy found is close to the true ground state.
        For $^{60}$Zn, achieving energy error less than half the gap requires the execution of the nuc-CR algorithm.
        To achieve a similar energy approximation quality for $^{28}$Si, one must perform the time-reversal of the correct and corrected bit strings.
        Finally, applying excitations to the bit strings allows to obtain nearly exact ground state energies for all the three isotopes.

        The results presented in Figures~\ref{fig:three_isotopes-bitstrings} and \ref{fig:three_isotopes-energy_errors} hint that different bit strings contribute differently to the ground state for each of the three isotopes.
        We verify this in Appendix~\ref{app:bitstrings-importance-gs} and observe that while the ground states of $^{48}$Ca and $^{60}$Zn are mostly concentrated on a few hundred bit strings, nearly all of them are important for $^{28}$Si.

\begin{figure*}
            \centering
            \includegraphics[width=0.715\linewidth]{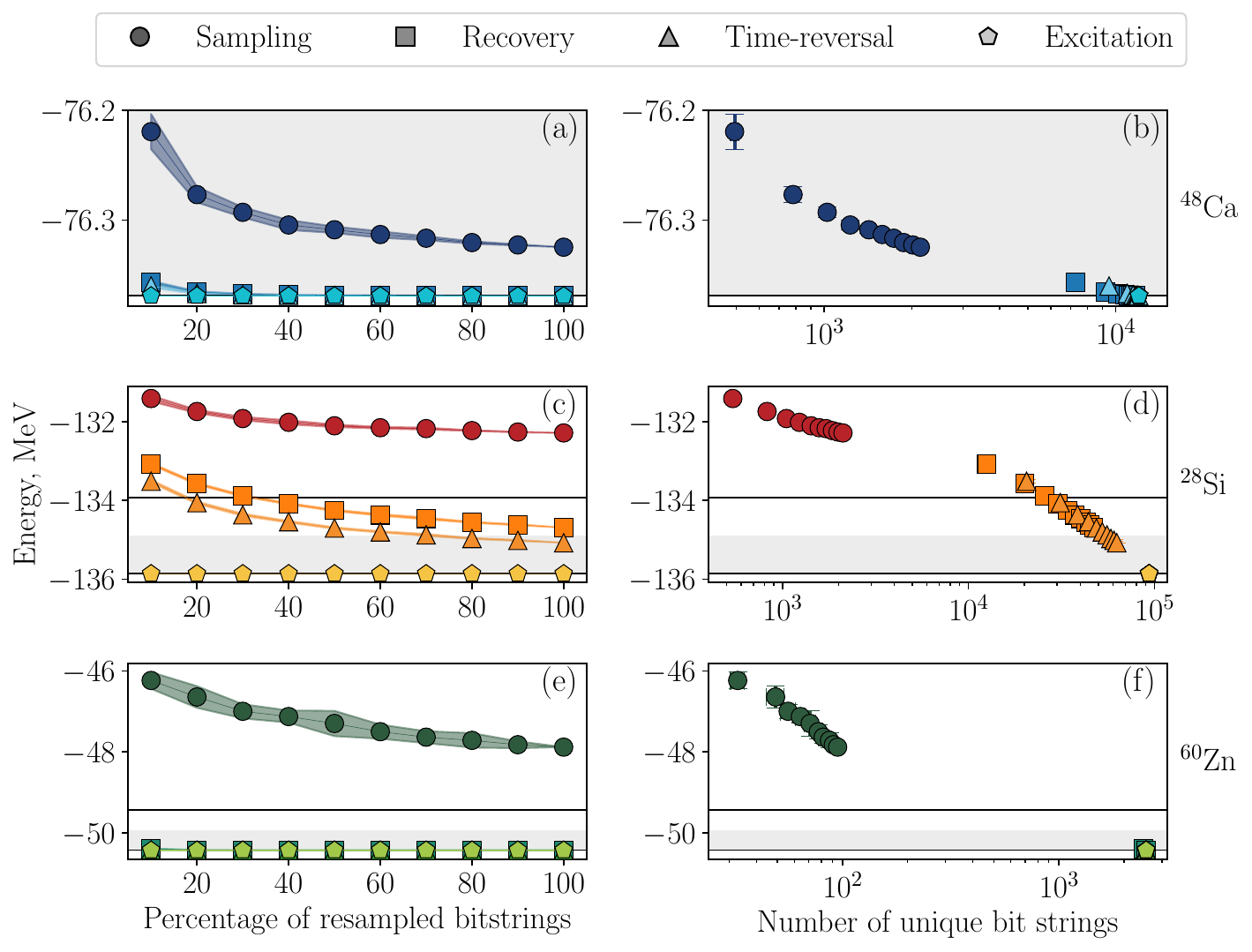}
            \caption{
                Approximate ground state energy $\tilde{E}_{\mathrm{gs}}$ for $^{48}$Ca, $^{28}$Si, and $^{60}$Zn.
                Panels (a), (c), and (e) show the energies obtained using 10\%, 20\%, ..., 100\%  of the (non-unique) determinants $\mathcal{S}$ randomly sub-sampled, with replacement, from the bit strings distribution measured on \texttt{ibm\_basquecountry}.
                Panels (b), (d), and (f) show $\tilde{E}_{\mathrm{gs}}$ as a function of subspace dimension corresponding to the sub-sampled bit strings.
                Different markers indicate different bit string acquisition stages with varying amounts of additional classical post-processing similarly to Figure~\ref{fig:three_isotopes-bitstrings-energy_errors}.
                Each data point corresponds to an average of 10 different sub-samplings instances.  Shaded areas and error bars show the standard deviation over these.
                The horizontal black lines indicate the exact energies of the ground $E_{\mathrm{gs}}$ and the first excited $E_{\mathrm{1ex}}$ states, and the gray area shows half of the energy gap used in \eqref{eq:gap_error}; 
                for $^{48}$Ca, $E_{\mathrm{1ex}} = -72.13753$ MeV is much higher than $\tilde{E}_{\mathrm{gs}}$ and therefore is not shown.
           }
            \label{fig:energies_vs_bitstrings}
        \end{figure*}

        To better understand the performance of the algorithm at different subspace sizes, in Figure~\ref{fig:energies_vs_bitstrings} we plot the approximate ground state energy $\tilde{E}_{\mathrm{gs}}$ against the number of determinants used for projection and diagonalization of $\Hnuc$.
        The figure shows the energies computed after randomly sub-sampling, without replacement, 10\%, 20\%, ..., 100\% of the $|\mathcal{S}| = 250,\!000$ bit strings measured on the quantum computer.
        These sub-sampled determinants, as before, are passed through configuration recovery, time-reversal, and excitations. 
        As we observed earlier, the sampled determinants for $^{48}$Ca are already enough to reach a small energy error.
        To obtain all bit strings in the correct symmetry sector for $^{60}$Zn, it is enough to use the nuc-CR algorithm with only 10\% of sampled determinants $\mathcal{S}$. 
        For $^{28}$Si, one can see that if we use only configuration recovery, we can reach energies slightly above half of the gap $\varepsilon_\Delta \approx 1.21$, although with the need of all the sampled bit strings.
        To descend closer to the ground state than to the first excited one, we need to apply the time-reversal, but using only up to about 70\% of $\mathcal{S}$.
        If one additionally applies excitations, even 10\% of measured bit strings are enough to generate all $M$-scheme basis states and therefore to achieve the true ground state energy.

\section{Discussion}
\label{sec:discussion}

    In this section, we discuss the results presented in this work, as well as the methods applied for obtaining them.
    Additionally, we outline potential directions for future work.


    \subsection{Quality of sampled bit strings}
    
        The noise in the quantum device corrupts the bit strings we sample from it.
        Indeed, the 250,000 shots we took produced relatively few bit strings with the correct quantum numbers $\nnuc$, $T_z$, $M$, and $\Pi$.
        This is especially true for $^{28}$Si, as we sample only $|\mathcal{C}| = 2118$ correct bit strings, which is about 2\% of the bit strings of the ground state of this isotope.
        However, the data obtained from the quantum processor is not hopelessly noisy, as we are able to repair it using the nuc-CR algorithm.
        As we show in Appendix~\ref{app:bitstrings-random}, the noisy bit strings are mostly only a few bits away from being correct.
        The rest of the bit strings needed for smaller energy errors can be obtained by applying the extension techniques such as the ones described in Section~\ref{sec:extension}.


    \subsection{Operations in the qDRIFT circuit}
    \label{sec:discussion-operations}

        The quality of sampled bit strings is primarily affected by the number and depth of noisy two-qubit gates. 
        As discussed in Section~\ref{sec:qdrift}, we reduced this depth by using low-depth construction of fermionic excitations $\exp\!\big(-i \theta (a^\dagger_\alpha a^\dagger_\beta a_\delta a_\gamma + \mathrm{H.c.}) \big)$. 
        Furthermore, one can filter out the sampled excitations that act trivially on a given reference state, and apply the symmetry shift technique \cite{loaiza2023reducing} for decreasing the error of qDRIFT (see Appendix~\ref{app:qdrift_improvements} for details). 
        The most significant remaining contributing factor to the gate depths is the distances between the qubits of the orbitals $\alpha$ and $\beta$, and $\gamma$ and $\delta$ (see Appendix~\ref{app:fermionic_gates} for details). Although not considered here, heuristic algorithms such as proposed in \cite{piccinelli2025quantum} can decrease these by permuting the orbital-to-qubit mapping (such as in Table~\ref{tab:o2q-ckpot}) thereby reducing the gate cost of the qDRIFT circuits \eqref{eq:qdrift}.
        
        The noise can also be reduced by exchanging the number of excitations $K$ for the number of the qDRIFT circuit randomizations $R$.
        Assuming that the initial state $\ket{\psi_0}$ contains $B$ bit strings, and noting that each excitation can \textit{at most} double the number of bit strings, one can sample \textit{no more than} $B R 2^K$ bit strings using SqDRIFT.
        Looking at this extremely rough upper bound, one sees that with a small $K$ we can find only a few ``low-excitation'' bit strings \cite{piccinelli2025quantum}.
        At the same time, as we show in Appendix~\ref{app:bitstrings-random}, having  many excitations $K$ may introduce too much noise so that the sampled bit strings become less distinguishable from randomly generated ones.
        Therefore, one needs to seek a balance between exploring wider spaces of Slater determinants (and achieving a better qDRIFT approximation error) and having less noise in the results.


    \subsection{Subtleties in configuration recovery}
    \label{sec:discussion-cr}

        The nuclear Hamiltonian \eqref{eq:H_nuc} has a number of symmetries, some of which are kept in the bit strings in \eqref{eq:gs_bitstrings} as quantum numbers $\nnuc$, $T_z$, $M$, and $\Pi$.
        While having many symmetries is commonly preferable, it complicates the CR algorithm.
        That is, as we discuss in Appendix~\ref{app:cr_algorithm}, imposing the constraints for preserving the desired quantum numbers when repairing the bit strings leads to solving a subset sum problem which is NP-complete.
        
        Luckily, it has a relatively simple linear programming formulation, even with a non-linear (but linearizable) constraint on the parity $\Pi$.
        However, one may consider basing the nuc-CR algorithm on other approaches such as dynamic programming or heuristic algorithms \cite{caprara2000approximation}.
        Another alternative may be transforming the incorrect bit strings such that they have the correct numbers of protons $\npro$ and neutrons $\nneu$ using, e.g., the original CR algorithm \cite{robledo2025chemistry}, and then projecting them onto the subspaces of desired symmetries \cite{nogaki2025symmetry}.

        The CR algorithm proposed in the original work~\cite{robledo2025chemistry} is iterative and self-consistent.
        That is, given an incorrect bit string, the algorithm flips its bits according to the average occupancies of the single-particle orbitals in the approximate ground state $|\tilde{\psi}_\mathrm{gs}\rangle$ at each iteration, and may be set to terminate if the occupancies did not change much between iterations.
        The earlier versions of our nuc-CR algorithm had these features as well.
        However, we observed that just finding the closest correct bit string to a given incorrect one generates more new bit strings.
        At the same time, it is not necessarily advantageous to just have a larger number of bit strings in the correct symmetry sector, as one is interested also in their importance and will eventually be limited in the size of diagonalization step that can be performed.
        Therefore, self-consistency may help to obtain more compact subspaces that produce similar ground state energies after projection and diagonalization.
        This can be tested on isotopes having more Slater determinants in their ground state, which we leave for the future work.


    \subsection{Future work}

        While we have demonstrated our workflow at scales up to 44 qubits, further nuclei of interest, e.g., for neutrinoless double-beta decay, can have ground states with support on more determinants. 
        $^{76}$Ge, for example, would also require just $44$ qubits within the model space of the JUN45 interaction \cite{honma2009new}, but its $M$-scheme dimension is 17,284,219.
        Studying such systems would require certain modifications of the machinery described in this work.
        
        First, additional improvement to the nuc-CR algorithm may be required to scale to larger active spaces and bias distributions further towards important bit strings.
        While solving the subset sum problem is a subroutine that seems hard to avoid, there may be approaches more efficient than linear programming used in this work. 

        Second, the slowest part of our workflow is computation of the projection of the Hamiltonian onto the sampled bit strings as in \eqref{eq:H_proj-eigenproblem}.
        For larger-scale computations, one may need a more efficient code, preferably suitable for high performance computing environments \cite{shirakawa2025closed}.

        Third, for more complex systems---such as $^{76}$Ge mentioned above---one may have to operate over millions or even billions of bit strings.
        To do that, one may need more efficient diagonalization algorithms based on subsampling \cite{robledo2025chemistry}, which may require adaptation to deal with the additional symmetry constraints of nuclear Hamiltonians.
        Additionally, one may need to discard the leastimportant bit strings, which requires the assessment of the impact of each bit string on the ground state energy \cite{holmes2016heat, barison2025quantum, stumpf2016importance, hoppe2022importance}.


\section{Conclusion}
\label{sec:conclusion}

    We have applied the SqDRIFT algorithm to effective nuclear shell model Hamiltonians.
    The algorithm was executed on the \texttt{ibm\_basquecountry} quantum processor for the Hamiltonians describing the three isotopes: $^{48}$Ca, $^{28}$Si, and $^{60}$Zn.
    We presented nuclear-shell model configuration recovery, an algorithm that uses symmetries of the nuclear Hamiltonian to recover determinants in the correct symmetry sector from those affected by noise on the quantum computer.
    This, combined with further post-processing using ``time-reversal'' and additional excitations, allowed us to improve the yield of relevant determinants from the quantum processor and recover accurate ground state energies.

    The machinery described in this work can also be used for finding the energies of excited states \cite{barison2025quantum,liepuoniute2025quantum}.
    Moreover, these energies can be obtained using the very same bit strings sampled for finding the ground state, provided that the excited state of interest has the same quantum numbers.
    In addition, more targeted SQD approaches can be considered for this task \cite{barison2025quantum, kanno2026quantum}.

    With our work, we hope to attract the attention of the nuclear physics community to the sample-based quantum diagonalization methods.
    Possible future work includes studying heavier and more complex isotopes, calculating nuclear matrix elements, optimizing our code for the classical part of the computations, and considering other ways of repairing incorrect bit strings.


\section{Code and data availability}
\label{sec:code_availability}

    The data and code used in this work are available upon reasonable request.
    The code uses the ffsim library \cite{sung2026ffsim} for representing and processing nuclear Hamiltonians $\Hnuc$.
    To compose quantum circuits and interact with quantum hardware, we use qiskit \cite{qiskit2024}.
    For error suppression and mitigation, we use Fire Opal \cite{fire_opal}.
    Linear programming problems are solved by means of PuLP \cite{pulp}.


\section{Acknowledgments}

   The authors acknowledge the support of the BasQ program through the NMEQ project. We are also thankful to D. Casanova, A. Carreras, and N. Vaquero Sabater (DIPC) for their precious feedback.

\bibliography{bibliography}
\bibliographystyle{apsrev4-2}

\onecolumngrid


\appendix

\section{$J$-to-$M$ transformation of TBMEs}
\label{app:J-to-M}

    As we mentioned in Section~\ref{sec:nuclear_shell_model}, in Hamiltonian \eqref{eq:H_nuc}, the TBMEs $\overline{v}_{\alpha \beta \gamma \delta}$ are assumed to be in the $M$-scheme, i.e., the two-nucleon states $\ket{\alpha\beta}$ and $\ket{\gamma\delta}$ are coupled to have a good total angular momentum projection $M$.
    There is also a more compact representation called the $J$-scheme, in which the two-nucleon states have a good total angular momentum $J$.
    In this form, the TBMEs are often used in software for nuclear shell model calculations \cite{shimizu2013nuclear, shimizu2019thick, bertsch2016finite}.
    Using the notation of \cite{suhonen2007nucleons} and defining $a \equiv (n^{(a)}, l^{(a)}, j^{(a)}, t_z^{(\alpha)})$, the $J$-scheme TBMEs are denoted as $\langle ab|V| cd\rangle_J$.
    Letting $\alpha \equiv (a, m^{(\alpha)})$, these $J$-scheme TBMEs $\langle ab|V| cd\rangle_J$  can be transformed to the $M$-scheme TBMEs $\overline{v}_{\alpha \beta \gamma \delta}$ as
    \begin{equation}
        \label{eq:J-to-M}
        \overline{v}_{\alpha \beta \gamma \delta} = \sum_{J,M} \frac{1}{\mathcal{N}_{ab}(J) \mathcal{N}_{cd}(J)} (j^{(a)} m^{(\alpha)} j^{(b)} m^{(\beta)}|JM) \; (j^{(c)} m^{(\gamma)} j^{(d)} m^{(\delta)}|JM) \; \langle ab|V| cd\rangle_J.
    \end{equation}
    Here, the summation is taken over the allowed total angular momenta $J$ $\in \big\{|j^{(a)} - j^{(b)}|, \ldots, j^{(a)} + j^{(b)} \big\} \cap \big\{ |j^{(c)} - j^{(d)}|, \ldots, j^{(c)} + j^{(d)} \big\}$, its projection satisfies $M = m^{(\alpha)} + m^{(\beta)} = m^{(\gamma)} + m^{(\delta)}$, the quantities $(j^{(a)} m^{(\alpha)} j^{(b)} m^{(\beta)}|JM)$ are the Clebsch-Gordan coefficients, and the normalization factors are
    \begin{equation*}
        \mathcal{N}_{ab}(J) = \sqrt{\frac{1 + \delta_{ab} (-1)^J}{1 + \delta_{ab}}}.
    \end{equation*}

    One often additionally assumes the isospin symmetry, when the two-nucleon states have also a well-defined isospin $T$.
    In this form, the TBMEs labeled by $JT$ are often listed in the literature (see, e.g., \cite{brown2006new}) and used in nuclear shell model calculation packages \cite{johnson2018bigstick, brown2014shell}.
    Denoting now $a \equiv (n^{(a)}, l^{(a)}, j^{(a)})$ and $\alpha \equiv (a, t_z^{(\alpha)}, m^{(\alpha)})$, the $J$-to-$M$ transformation formula is 
    \begin{multline}
        \label{eq:JT-to-M}
        \overline{v}_{\alpha \beta \gamma \delta} = \sum_{J,M,T,T_z} \frac{1}{\mathcal{N}_{ab}(JT) \mathcal{N}_{cd}(JT)} (j^{(a)} m^{(\alpha)} j^{(b)} m^{(\beta)}|JM) \; (\frac12 t_z^{(\alpha)} \frac12 t_z^{(\beta)}|T T_z) \\ 
        \times (j^{(c)} m^{(\gamma)} j^{(d)} m^{(\delta)}|JM) \; (\frac12 t_z^{(\gamma)} \frac12 t_z^{(\delta)}|T T_z) \; \langle ab|V| cd\rangle_{JT},
    \end{multline}
    where $T_z = t_z^{(\alpha)} + t_z^{(\beta)} = t_z^{(\gamma)} + t_z^{(\delta)}$ and 
    \begin{equation*}
        \mathcal{N}_{ab}(JT) = \sqrt{\frac{1 - \delta_{ab} (-1)^{J + T}}{1 + \delta_{ab}}}.
    \end{equation*}


\section{Fermionic excitation circuits}
\label{app:fermionic_gates}

    In Section~\ref{sec:qdrift}, we described the qDRIFT algorithm which approximates the Hamiltonian evolution as a product of $e^{-i \theta E_l}$, where $\theta \in \mathbb{R}$, and $E_{l} = a^\dagger_\alpha a_\alpha$ or $E_{l} = a^\dagger_\alpha a^\dagger_\beta a_\delta a_\gamma + a^\dagger_\gamma a^\dagger_\delta a_\beta a_\alpha$ are picked from the nuclear Hamiltonian \eqref{eq:H_nuc}.
    Composing the qDRIFT circuit \eqref{eq:qdrift} of operators of this form was proposed in \cite{piccinelli2025quantum} for preparing the Krylov states \eqref{eq:krylov_states}. 
    The authors first used the Jordan-Wigner (JW) transformation to map the Hamiltonian in question into a sum of Pauli strings, grouped them according to the order of the operators in \eqref{eq:qdrift}, and simulated each Pauli string consequently.
    When acting on a state $\ket{\psi_0}$, grouping the Pauli strings in this manner allows us to preserve the quantum numbers $\nnuc$, $T_z$, $M$, and $\Pi$ in our case.

    The application of the JW transformation to the operators $E_{l}$ in \eqref{eq:E_operators} may produce up to eight Pauli strings for each operator.
    Each of these Pauli strings can be simulated on a quantum computer efficiently using, e.g., the so-called staircase algorithm requiring two chains of CNOT gates connecting the qubits the Pauli string acts on \cite{mansky2023decomposition, carreras2026limitations}.
    This therefore results in up to 16 staircases of two-qubit gates for each operator $E_{l}$.

\begin{figure}[h]
        \centering
        \includegraphics[width=1\linewidth]{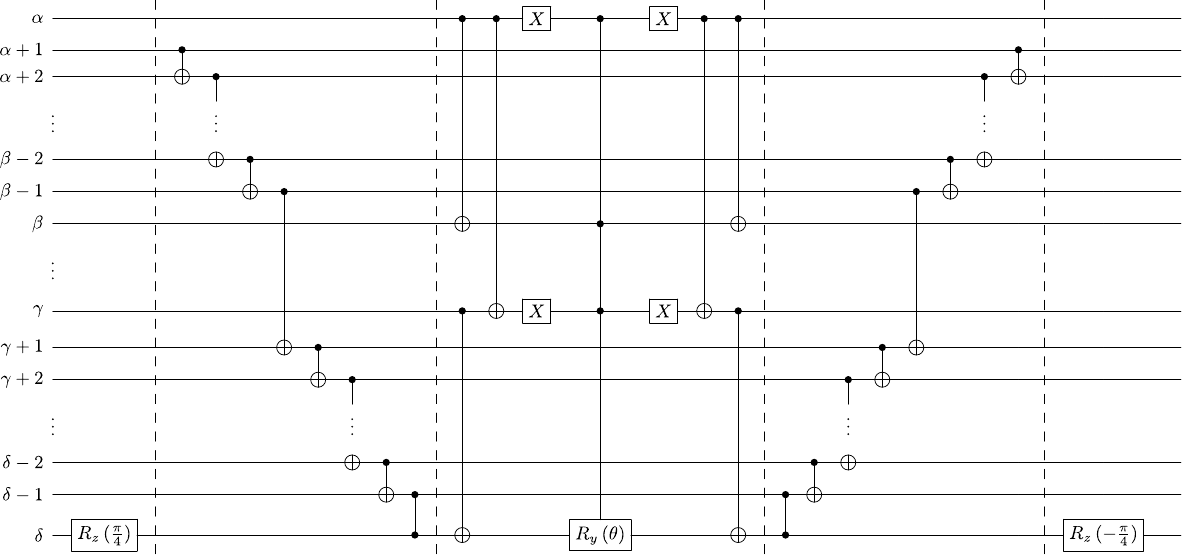}
        \caption{
            Quantum circuit implementing a fermionic double excitation $\exp(- i \theta (a^\dagger_\alpha a^\dagger_\beta a_\delta a_\gamma + a^\dagger_\gamma a^\dagger_\delta a_\beta a_\alpha))$ for $\alpha < \beta < \gamma < \delta$. 
            Adapted from \cite{yordanov2020efficient}, where one can also find an efficient decomposition of the 3-controlled $y$-rotation used in this work.
        }
        \label{fig:fde_circuit}
    \end{figure}

    However $e^{-i \theta E_l}$ are nothing but fermionic excitation operators for which efficient quantum circuits are known \cite{yordanov2020efficient}.
    For instance, in Figure~\ref{fig:fde_circuit} shown is the circuit for the case when all the orbital indices are distinct, so that $\exp(- i \theta (a^\dagger_\alpha a^\dagger_\beta a_\delta a_\gamma + a^\dagger_\gamma a^\dagger_\delta a_\beta a_\alpha))$ is a fermionic double excitation.
    This operation effectively performs a rotation by the angle $\theta$ in the subspace $\{\ket{0_\alpha 0_\beta 1_\gamma1_\delta }, \ket{1_\alpha 1_\beta 0_\gamma 0_\delta }\}$, which is done by the central block of the circuit in the figure.
    This block, including the 3-controlled $y$-rotation, decomposes into 13 CNOTs, see \cite{yordanov2020efficient}.
    To account for the anticommutation relations of the fermionic creation and annihilation operators $a^\dagger$ and $a$, two staircases of CNOTs are used.
    These staircases accumulate the parity of the qubits from $(\alpha + 1)$ to $(\beta - 1)$, and from $(\gamma + 1)$ to $(\delta - 1)$.\footnote{This holds only for $\alpha < \beta < \gamma < \delta$. Generally, one needs to check the parity of the qubits in the disjunctive union of $\{\alpha + 1, \ldots, \beta - 1\}$ and $\{\gamma + 1, \ldots, \delta - 1\}$.}
    This parity is stored in the qubit $(\delta - 1)$, which is then transferred to the qubit $\delta$ via a $\mathrm{CZ}$ gate to change the rotation sign of the 3-controlled $y$-rotation.
    Lastly, in \cite{yordanov2020efficient} the circuits are derived to implement $\exp(\theta (a^\dagger_\alpha a^\dagger_\beta a_\delta a_\gamma - a^\dagger_\gamma a^\dagger_\delta a_\beta a_\alpha))$, which is turned to the desired form $\exp(- i \theta (a^\dagger_\alpha a^\dagger_\beta a_\delta a_\gamma + a^\dagger_\gamma a^\dagger_\delta a_\beta a_\alpha))$ by conjugating with $\exp(-i \frac{\pi}{4} a^\dagger_\delta a_\delta)$; this is done by introducing the $R_z(\pm\pi/4)$ gates. 

\begin{figure}[h]
        \centering
        \includegraphics[width=0.65\linewidth]{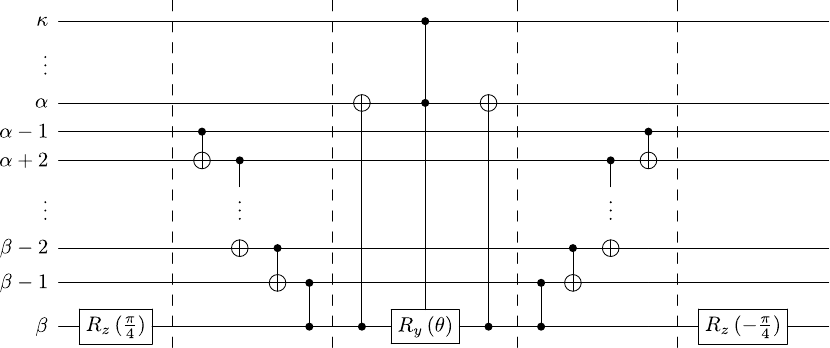}
        \caption{
            Quantum circuit implementing a controlled fermionic single excitation $\exp(-i \theta n_\kappa (a^\dagger_\alpha a_\beta + a^\dagger_\beta a_\alpha))$ for $\kappa < \alpha < \beta$. 
            Synthesized using techniques from \cite{yordanov2020efficient, ramoa2025reducing}.
            For an efficient decomposition of the 2-controlled $y$-rotation, see \cite{ramoa2025reducing}.
        }
        \label{fig:cse_circuit}
    \end{figure}

    In Hamiltonian \eqref{eq:H_nuc}, there may be terms with pairs of $a^\dagger$ and $a$ sharing the same single-particle orbital $\kappa$, i.e.,
    \begin{align*}
        a^\dagger_\alpha a^\dagger_\kappa a_\kappa a_\gamma + a^\dagger_\gamma a^\dagger_\kappa a_\kappa a_\alpha &= n_\kappa (a^\dagger_\alpha a_\gamma + a^\dagger_\gamma a_\alpha), \\
        a^\dagger_\kappa a^\dagger_\beta a_\delta a_\kappa + a^\dagger_\kappa a^\dagger_\delta a_\beta a_\kappa &= n_\alpha (a^\dagger_\delta a_\beta + a^\dagger_\beta a_\delta), \\
        a^\dagger_\alpha a^\dagger_\kappa a_\delta a_\kappa + a^\dagger_\kappa a^\dagger_\delta a_\kappa a_\alpha &= -n_\kappa (a^\dagger_\alpha a_\delta + a^\dagger_\delta a_\alpha), \\
        a^\dagger_\kappa a^\dagger_\beta a_\kappa a_\gamma + a^\dagger_\gamma a^\dagger_\kappa a_\beta a_\kappa &= -n_\kappa (a^\dagger_\beta a_\gamma + a^\dagger_\gamma a_\beta),
    \end{align*}
    where $n_\kappa \equiv a^\dagger_\kappa a_\kappa$ is the particle number operator.
    The exponentiation of these terms gives a controlled fermionic single excitation of the form $\exp( -i\theta n_\kappa (a^\dagger_\alpha a_\beta + a^\dagger_\beta a_\alpha))$ which effectively acts in the subspace $\{\ket{1_\kappa 0_\alpha 1_\beta}, \ket{1_\kappa 1_\alpha 0_\beta}\}$.
    A circuit implementing this operation is shown in Figure~\ref{fig:cse_circuit}.
    Similarly to the double excitation, the parity between $\alpha$ and $\beta$ is computed using CNOT staircases.\footnote{Care must be taken when $\alpha < \kappa < \beta$. In this case, since the gate works only if the qubit $\kappa$ is in the state $\ket{1}$, the accumulated parity is automatically incremented by 1, so one should additionally multiply $\theta$ by $-1$.}
    The rotation by the angle $\theta$ in the desired subspace involves a 2-controlled $y$-rotation gate.
    The whole rotation block can be decomposed into 2 CNOTs and 4 CZs, see \cite{ramoa2025reducing}. 

    Among the terms in Hamiltonian \eqref{eq:H_nuc} we may also have the ones with $a^\dagger$ and $a$ sharing two pairs of orbitals, i.e.,  $(a^\dagger_\alpha a^\dagger_\beta a_\beta a_\alpha + a^\dagger_\alpha a^\dagger_\beta a_\beta a_\alpha) = n_\alpha n_\beta$.
    The operation $\exp(-i\theta n_\alpha n_\beta)$ is just a controlled $z$-rotation by the angle $\theta$.
    Finally, the exponentials of the single-body terms in \eqref{eq:H_nuc}, $\exp(- i\theta n_\alpha)$, are single-qubit $z$-rotations.

\begin{figure}[h]
        \centering
        \includegraphics[width=0.275\linewidth]{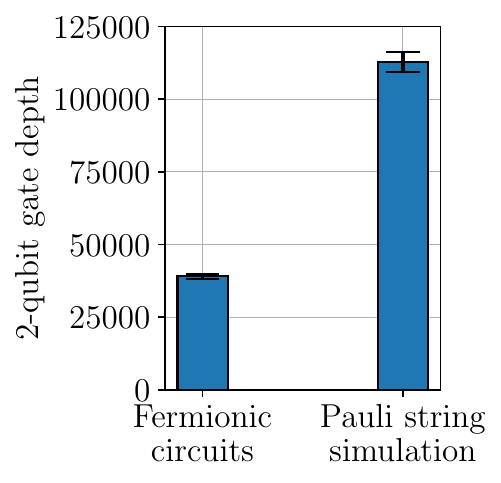}
        \caption{
            Two-qubit gate depths of the qDRIFT circuit constructed using fermionic excitation circuits (left bar), or applying the Jordan-Wigner transformation and simulating the Pauli strings (right bar).
            The qDRIFT sequence contains $K=1000$ operators sampled from $\Hnuc$ describing the $^{28}$Si isotope.
            The circuits are transpiled for the \texttt{ibm\_basquecountry} quantum machine.
            The error bars indicate the standard deviation obtained by averaging over $R=50$ qDRIFT circuit randomizations.
        }
        \label{fig:2q-depths}
    \end{figure}

    To show the advantage of the circuits described above, we first construct a qDRIFT sequence of $K=1000$ operations sampled from the Hamiltonian of the $^{28}$Si isotope.
    Then, we turn this sequence into two quantum circuits: One by using the fermionic excitation circuits, and the other by applying the Jordan-Wigner transformation to $a^\dagger$ and $a$ and simulating the resultant Pauli strings. 
    In Figure~\ref{fig:2q-depths}, we show the two-qubit gate depths for both circuits transpiled for the \texttt{ibm\_basquecountry} quantum computer.
    As one can see, the qDRIFT circuits composed of fermionic excitations are more than $2.8$ times shallower. 

\section{Improving the performance of qDRIFT}
\label{app:qdrift_improvements}
        
        In this section, we describe the techniques we use for improving the performance of the qDRIFT algorithm.
        
        \subsection{Removing trivially acting operators}

            After sampling the operators for the qDRIFT circuit \eqref{eq:qdrift}, some of them may do nothing to the initial state.
            That is, for instance, the excitations $E_{\alpha \beta \gamma \delta}(\theta) =  \exp(- i \theta (a^\dagger_\alpha a^\dagger_\beta a_\delta a_\gamma + a^\dagger_\gamma a^\dagger_\delta a_\beta a_\alpha))$ act trivially outside of the subspace spanned by $\{\ket{0_\alpha 0_\beta 1_\gamma1_\delta }, \ket{1_\alpha 1_\beta 0_\gamma 0_\delta }\}$.
            Therefore, at the very beginning of the qDRIFT circuit, this operator may be very likely to produce no excitations.
            Since each gate introduces an error to the quantum state prepared on real hardware, it is better to remove such operations from the circuit.
            We do this by tracking the orbitals ``touched'' by the excitations applied.  
            
            As an example, consider the initial state to be $\ket{\psi_0} = \ket{1_{1} 1_{2} 0_{3} 0_{4} 0_{5} 0_{6} }$, where the subscripts index the single-particle orbitals $\alpha$.
            In $\ket{\psi_0}$, the occupied orbitals 1 and 2 are considered to be active.
            If the excitation $E_{3,4,5,6}$ is sampled for the qDRIFT circuit, it would not change this state, so it is discarded from the sequence.
            In contrast, $E_{1,2,3,4}$ would excite nucleons from orbitals 1 and 2 to 3 and 4, so $\ket{\psi_0}$ becomes a superposition of $\ket{1_{1} 1_{2} 0_{3} 0_{4} 0_{5} 0_{6} }$ and $\ket{0_{1} 0_{2} 1_{3} 1_{4} 0_{5} 0_{6} }$.
            Orbitals 3 and 4 have been ``touched'' by an excitation, so now we also consider them to be active together with 1 and 2.
            Now the excitation $E_{3,4,5,6}$ ---which did nothing before--- may potentially act non-trivially \textit{after} $E_{1,2,3,4}$ has been applied.
            
            This way, we track the orbitals ``touched'' by sampled excitations until they all become active,
            and the excitations that are guaranteed to act trivially are removed.

        
        \subsection{Symmetry shift}

            The error between the exact evolution $e^{-i \Hnuc t}$ and its qDRIFT approximation \eqref{eq:qdrift} is known to scale as $O(\lambda^2 t^2/K)$, where $\lambda = \sum_l |e_l|$ is the sum of the magnitudes of the coefficients of the Hamiltonian's terms \cite{campbell2019random}.
            The quantity $\lambda$ is often called the 1-norm of the Hamiltonian $H$, $||H||_1$.
            This norm is important for some quantum algorithms, as their performance depends on it \cite{childs2012hamiltonian, low2019hamiltonian, lee2021even, koridon2021orbital, granet2024hamiltonian}.
            One approach to decreasing $\lambda$ is the symmetry shift method proposed in \cite{loaiza2023reducing}.

            Suppose that $S$ is a symmetry of $H$, i.e., $[H, S] = 0$.
            Assume also that $S\ket{\psi_0} = s\ket{\psi_0}$, i.e., the initial state $\ket{\psi_0}$ we want to evolve under $H$ is an eigenstate of the symmetry $S$.
            Then one has $e^{-i t H} \ket{\psi_0} = e^{-its} e^{-it(H - S)}\ket{\psi_0}$.
            That is, the evolution of $\ket{\psi_0}$ under $H$ is equivalent to that under $(H - S)$ up to a phase $e^{-its}$.
            Therefore, one may try to find a symmetry $S$ that minimizes $||H - S||_1$.

            For our nuclear Hamiltonian $\Hnuc$, we chose the symmetry to be $S(c_1, c_2) = c_1 \hat{n}_{\mathrm{nuc}} + c_2 \hat{n}^2_{\mathrm{nuc}}$, where $\hat{n}_{\mathrm{nuc}} = \sum_{\alpha=1}^{\norb} a^\dagger_\alpha a_\alpha$ is the total particle number operator.
            To optimize the coefficients $c_1$ and $c_2$, we use the linear program described in \cite{loaiza2023reducing}.


\section{Details of the configuration recovery algorithm}
\label{app:cr_algorithm}

    To repair the incorrect bit strings sampled using SqDRIFT, we apply a configuration recovery (CR) algorithm. 
    In this section, we first outline the original algorithm proposed in \cite{robledo2025chemistry} for quantum chemistry problems, and then we describe in detail our nuc-CR algorithm for nuclear Hamiltonians.

    \subsection{CR algorithm for electronic Hamiltonians}

        When applying SQD to quantum chemistry Hamiltonians $H_{\mathrm{el}}$, one commonly cares about two quantum numbers preserved in a sampled bit string $s \in \mathcal{S}$: the number of electrons $n_\mathrm{el}$ (analogous to $\nnuc$), and the total spin projection $S_z$ (analogous to $T_z$).
        Therefore, verifying whether a bit string $s$ is correct reduces to checking if $s$ corresponds to a Slater determinant with the correct number of spin-up $n_{\uparrow}$ and spin-down $n_{\downarrow}$ electrons occupying the corresponding spin-orbitals $\alpha$.
        Thus one can straightforwardly split the sampled bit strings $\mathcal{S}$ into the correct $\mathcal{C}$ and incorrect $\mathcal{I}$ bit strings, $\mathcal{S} = \mathcal{C} \cup \mathcal{I}$.

        For repairing incorrect bit strings $\mathcal{I}$ using the information about the correct ones $\mathcal{C}$, the work \cite{robledo2025chemistry} proposed a CR algorithm.
        From a high-level, simplified perspective, it works as follows.
        Starting with an empty set of recovered bit strings $\mathcal{R} = \emptyset$, the algorithm performs iteratively the following steps.
        \begin{enumerate}
            \item Project $H_{\mathrm{el}}$ onto $\mathcal{B} \equiv \mathcal{C} \cup \mathcal{R}$ as $\tilde{H}_{\mathrm{el}} = \sum_{x,y \in \mathcal{B}} \langle x | H_{\mathrm{el}} | y\rangle \ketbra{x}{y}$ and find the approximate ground state $\ket{\tilde{\psi}_{\mathrm{gs}}} = \sum_{b \in \mathcal{B}} g_b \ket{b}$ by solving $\tilde{H}_{\mathrm{el}} \ket{\tilde{\psi}_{\mathrm{gs}}} = \tilde{E}_{\mathrm{gs}} \ket{\tilde{\psi}_{\mathrm{gs}}}$.
            \item From $\ket{\tilde{\psi}_{\mathrm{gs}}}$, determine the average occupancy $\overline{n}_\alpha = \langle\tilde{\psi}_{\mathrm{gs}}| a^\dagger_\alpha a_\alpha | \tilde{\psi}_{\mathrm{gs}}\rangle$ of each spin-orbital $\alpha$.
            \item For each incorrect $i \in \mathcal{I}$, generate a recovered bit string $r(i)$ as follows.
            \begin{itemize}
                \item Find the number of spin-up orbitals $n_{\uparrow}^{(i)}$ occupied in $i$.
                \item If $n_{\uparrow}^{(i)}  > n_{\uparrow}$ (or $n_{\uparrow}^{(i)}  < n_{\uparrow}$), flip $|n_{\uparrow}^{(i)}  - n_{\uparrow}|$ bits which correspond to the spin-up orbitals $\alpha$ and have the values $i_\alpha = 1$ (or $i_\alpha = 0$) with probabilities weighted by $|i_\alpha - \overline{n}_\alpha|$.
                \item Do the same for the bits of the spin-down orbitals $n_{\downarrow}^{(i)}$.
            \end{itemize}
            \item Redefine the set of recovered bit strings $\mathcal{R} = \{r\}_r$.
        \end{enumerate}
        The algorithm therefore transforms an incorrect bit string $i$ to a bit string $r(i)$ having the correct numbers of spin-up $n_{\uparrow}$ and spin-down $n_{\downarrow}$ electrons.
        This is done by probabilistically flipping certain bits, depending on whether $i$ contains too many (or too few) spin-up and spin-down orbitals occupied.
        The probability of flipping is determined from the approximate ground state obtained at each iteration.
        This CR algorithm is self-consistent in the sense that the average orbital occupancies $\{\overline{n}_\alpha \}_{\alpha=1}^{\norb}$ converge to some values, so one can impose a termination criterion based on the change of the occupancies between iterations.
        
        We note that above presented is a simplified version of the original CR algorithm.
        The original CR algorithm has several additional features, such as constructing (possibly overlapping) batches of bit strings, diagonalizing the Hamiltonian within each batch, and selecting the lowest-energy result.
        The batches are formed according to the observed frequencies of the measured bit strings.
        Batching may be necessary when the number of obtained configurations is too large for the Hamiltonian projection and diagonalization.
        Nonetheless, the core routine of the algorithm is probabilistically flipping the bits of bit strings to restore the correct particle numbers.


    \subsection{CR algorithm for nuclear Hamiltonians}

        To better understand why the CR algorithm for electronic Hamiltonians is not directly applicable to nuclear  Hamiltonians of the form \eqref{eq:H_nuc}, let us consider a simple example.
        Suppose that we have applied the SqDRIFT algorithm to $\Hnuc$ describing $^{8}$Be within the model space of the Cohen-Kurath interactions, which we mentioned in Sections~\ref{sec:orbital-to-qubit} and \ref{sec:extension}.
        Recall that this isotope has $\npro = 2$ valence protons and $\nneu = 2$ valence neutrons giving the particle number $\nnuc = 4$ and the total isospin projection $T_z = 0$; the total angular momentum is known to be $J^\Pi=0^+$ which implies its projection $M=0$.

        Now consider the orbital-to-qubit mapping in Table~\ref{tab:o2q-ckpot} in the main text.
        Assume that among the sampled bit strings $\mathcal{S}$ we have $s = 11 \svert 0000 \dsvert 10 \svert 1000$.
        One can see that this bit string has the correct $\nnuc$, $T_z$, and $\Pi$, but $M=-2$.
        That is, while being a valid bit string for an electronic Hamiltonian (with changing $\nnuc$ to $n_{\mathrm{el}}$ and $T_z$ to $S_z$), it is not correct for a nuclear Hamiltonian.
        To repair such a string, we have to move the 1's (i.e., swap bits) such that we obtain a desired $M$ and preserve $T_z$ and $\Pi$.
        The main obstacle here is that finding the bits which can be swapped involves solving a subset sum problem.
        Indeed, having the quantum numbers $\{t_z^{(\alpha)}\}_{\alpha = 1}^{\norb}$ and $\{m^{(\alpha)}\}_{\alpha = 1}^{\norb}$ (see Table~\ref{tab:o2q-ckpot} in the main text), one wants to pick $\nnuc = \npro + \nneu$ elements from both sets such that they sum to a desired $T_z$ and $M$.
        The same applies to the orbital angular momenta $\{l^{(\alpha)}\}_{\alpha = 1}^{\norb}$, as the picked elements must also give the correct parity $\Pi$. 

        As mentioned in the main text, we can perform configuration recovery by solving  the optimization problem \eqref{eq:recovered_bitstring}. 
        Recall that it seeks for a bit string $s$ with bits $s_\alpha \in \{0, 1\}$ that has correct quantum numbers $\nnuc$, $M$, $T_z$, $\Pi$, and has a minimum distance to a given incorrect bit string $i$.
        One can turn \eqref{eq:recovered_bitstring} into a linear programming problem by linearizing the parity constraint $(-1)^{\sum_{\alpha=1}^{\norb} s_\alpha l^{(\alpha)}} = \Pi$.
        First, denoting $L \equiv \sum_{\alpha=1}^{\norb} s_\alpha l^{(\alpha)}$ and defining $P \equiv \frac{1 - \Pi}{2}$, the constraint can be rewritten as $L = P \; (\bmod\; 2)$. 
        This means that the difference between $L$ and $P$ is an even number, or $L - P  = 2k$ with $k$ being a non-negative integer. 
        Therefore, considering $k$ to be an auxiliary variable, \eqref{eq:recovered_bitstring} can be rewritten as a linear program: 
        \begin{align}
            \label{eq:cr-lp}
            \begin{split}
                &r(i) = \arg\max_{s} \sum_{\alpha=1}^{\norb} s_\alpha  i_{\alpha} \\
                &\text{s.t. } \sum_{\alpha=1}^{\norb} s_\alpha = \nnuc, \qquad \sum_{\alpha=1}^{\norb} s_\alpha t_z^{(\alpha)} = T_z, \qquad  \sum_{\alpha=1}^{\norb} s_\alpha m^{(\alpha)} = M,\\
                &\phantom{s.t.\;\,}\sum_{\alpha=1}^{\norb} s_\alpha l^{(\alpha)} - 2k = \frac{1 - \Pi}{2}, \qquad k \geqslant 0.
            \end{split}
        \end{align}
        Now the optimization problem and its constraints are linear in the binary variables $s_\alpha$ and the integer variable $k$.
        It can be solved using packages such as PuLP \cite{pulp}.

        In contrast to the original CR algorithm, ours is much simpler as it is not iterative, nor does it have a self-consistency property. 
        As we mentioned in Section~\ref{sec:discussion-cr}, an earlier version of our algorithm had these features.
        Namely, alongside with maximizing the overlap between the given incorrect bit string $i$ and the probe string $s$ ---to involve some utilization of the correct bit strings $\mathcal{C}$--- it also maximized the sum of the average occupancies $\sum_{\alpha=1}^{\nnuc} s_\alpha \overline{n}_\alpha$ obtained at each iteration.
        However, this self-consistent iterative algorithm showed to produce fewer bit strings than the simpler algorithm based on just solving \eqref{eq:cr-lp}.
        
        We note that having a larger number of correct bit strings is not necessarily an advantage.
        Indeed, one is also interested in the compactness of the subspace used for the projection and diagonalization.
        That is, having a smaller number of important bit strings is more preferable than having many unimportant ones (provided that the total importance is the same).
        Therefore, the self-consistency of the original CR algorithm \textit{may} lead to achieving more compact subspaces.
        In principle, one can alternate between the original CR (restores $\nnuc$ and $T_z$) and nuc-CR (restores $M$ and $\Pi$) algorithms, or synthesize a hybrid algorithm such as an earlier version of our nuc-CR mentioned above.
        We leave this direction for the future work, as it is better to investigate this for more complex isotopes with a greater number of Slater determinants in the ground state.


\section{Importance of Slater determinants in the exact ground states}
\label{app:bitstrings-importance-gs}

\begin{figure}[h]
        \centering
        \subfloat[]{
            \includegraphics[width=0.4\linewidth]{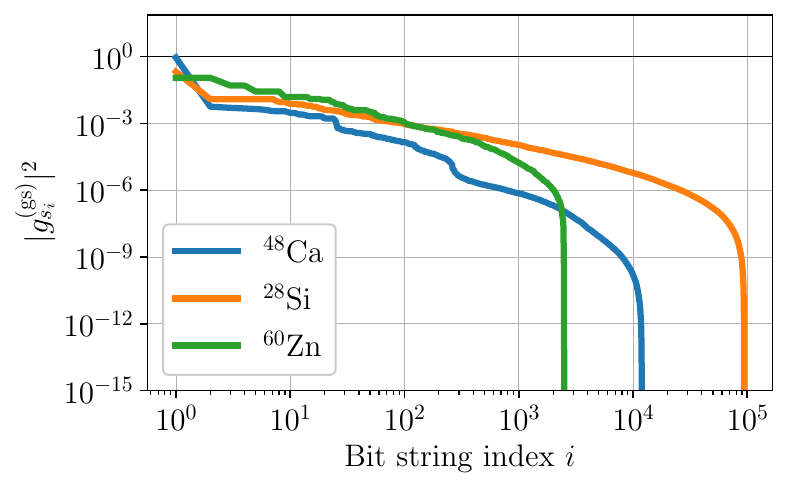}
        }
        \hspace{1.5cm}
        \subfloat[]{
            \includegraphics[width=0.4\linewidth]{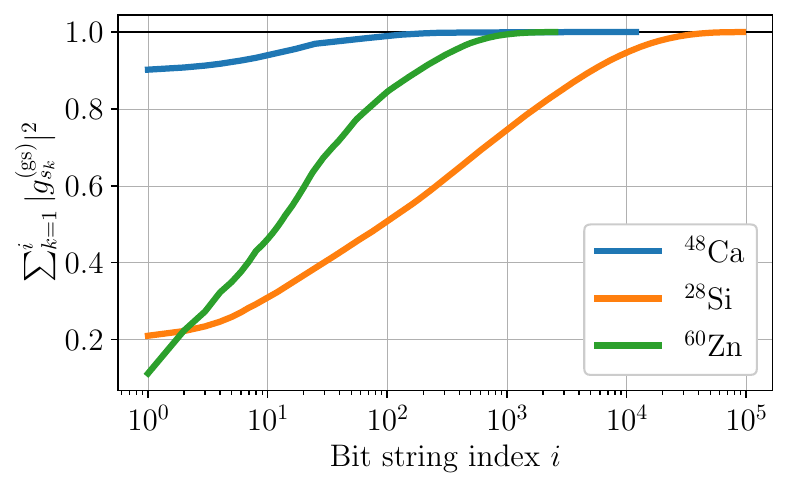}
        }
        \caption{
            (a) Squared magnitudes of the coefficients $g_s^{\mathrm{(gs)}}$ of bit strings $s$ of the ground state \eqref{eq:gs_bitstrings} for $^{48}$Ca, $^{28}$Si, and $^{60}$Zn.
            (b) Cumulative sum of $|g_s^{\mathrm{(gs)}}|^2$.
            In both plots, the bit strings $s$ are sorted in descending order of $|g_s^{\mathrm{(gs)}}|$.
        }
        \label{fig:bitstrings-probs}
    \end{figure}

    In our experiments with SqDRIFT, we observed that the three isotopes $^{48}$Ca, $^{28}$Si, and $^{60}$Zn  require different post-processing overheads for achieving lower approximate ground state energies.
    The reason for this is the sparsity of the ground states, i.e., the number of Slater determinants having high importance.
    We illustrate this in Figure~\ref{fig:bitstrings-probs}, where shown are the (squared) magnitudes of $g_s^{\mathrm{(gs)}}$ of each bit string $s$ of the exact ground state \eqref{eq:gs_bitstrings}.
    
    In the figure, one can see that the most significant bit string for $^{48}$Ca has the importance of more than 0.9. 
    This is in contrast with $^{28}$Si and $^{60}$Zn, for which many bit strings contribute non-negligibly to the ground state.
    Similar properties have recently been reported for the Heisenberg and Hubbard Models \cite{gaberle2026critical}.
    Therefore, one may seek to make the ground state more sparse by, e.g., (unitarily) transforming the Hamiltonian, or using more sophisticated techniques \cite{xu2026filter}. 
    
\section{Analysis of the quality of sampled bit strings}
\label{app:bitstrings-random}

    In our work, the SqDRIFT algorithm did not produce many correct bit strings for each of the studied isotopes.
    Indeed, for, e.g., $^{28}$Si among the $\mathcal{S} = 250000$ sampled bit strings only $\mathcal{C} = 2118$ unique ones out of 93710 have the desired quantum numbers (see Tables~\ref{tab:isotopes} and \ref{tab:three_isotopes-bitstrings}). 
    Additional correct Slater determinants can be generated from the incorrect ones using the nuc-CR algorithm.
    In principle, however, as long as the linear programming problem \eqref{eq:cr-lp} is feasible, we can ``repair'' \textit{any} given bit string. 
    This raises a question whether the incorrect bit strings $\mathcal{I}$ obtained from a quantum computer are useful.

\begin{figure}[h]
        \centering
        \subfloat[\label{fig:random_bitstrings-hamming-a}]{
            \includegraphics[width=0.3275\linewidth]{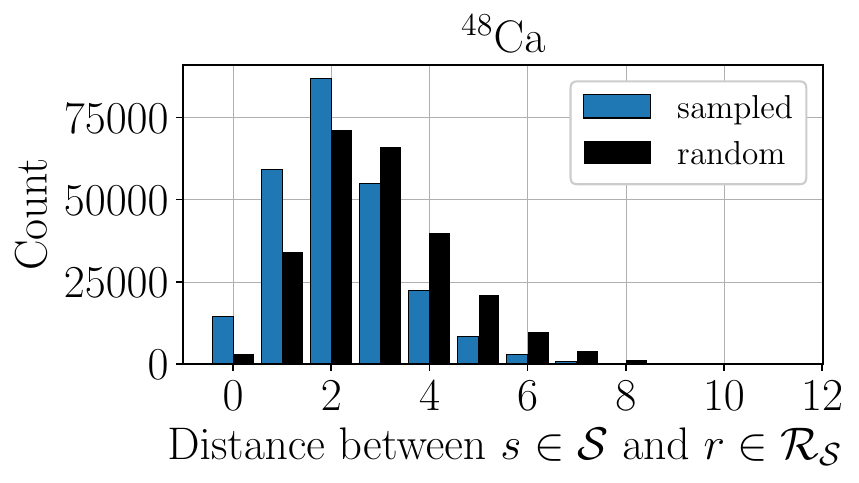}
        }
        \subfloat[\label{fig:random_bitstrings-hamming-b}]{
            \includegraphics[width=0.3275\linewidth]{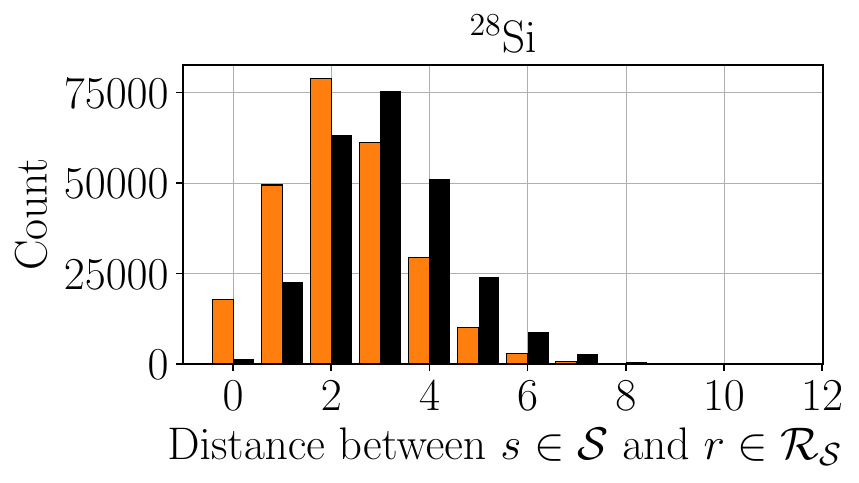}
        }
        \subfloat[\label{fig:random_bitstrings-hamming-c}]{
            \includegraphics[width=0.3275\linewidth]{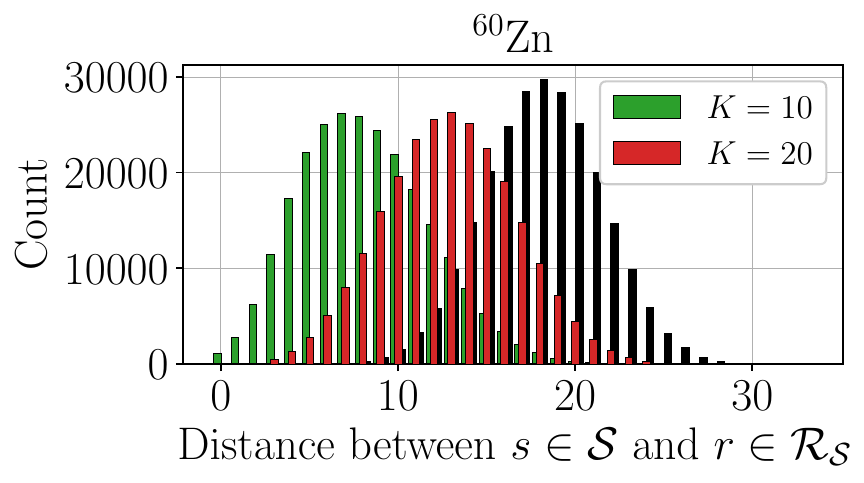}
        }
        \caption{
            Hamming distances between all the sampled $s \in \mathcal{S}$ and recovered $r \in \mathcal{R}_{\mathcal{S}}$ bit strings for (a) $^{48}$Ca, (b) $^{28}$Si, and (c) $^{60}$Zn.
            For the latter isotope, the distances are shown for the cases of $K=10$ (green) and $K=20$ (red) operations in the qDRIFT circuit \eqref{eq:qdrift}.
            Black bars indicate the data for 250000 bit strings generated randomly and uniformly.
        }
        \label{fig:random_bitstrings-hamming}
    \end{figure}

    In Figure~\ref{fig:random_bitstrings-hamming}, we show the Hamming distances (i.e., the number of differing bits) between the sampled $s \in \mathcal{S}$ bit strings and the bit strings $r \in \mathcal{R}_{\mathcal{S}}$ obtained after applying the nuc-CR algorithm to \textit{all} bit strings in $\mathcal{S}$.
    Additionally, we show these distances for randomly generated bit strings.
    As can be seen, the quantumly sampled bit strings demand flipping fewer bits for repairing.\footnote{Note that here we apply nuc-CR to \textit{all} the sampled bit strings $\mathcal{S}$, which contains \textit{both} the correct $\mathcal{C}$ and incorrect $\mathcal{I}$ strings. As nuc-CR does not change the correct strings $\mathcal{C}$, we observe a few counts of zero distances between $s \in \mathcal{S}$ and $r \in \mathcal{R}_{\mathcal{S}}$.}
    This is especially evident for $^{60}$Zn, for which the incorrect bit strings from quantum hardware are ---on average--- about 8 bits away from being correct, while one needs to flip about 18 bits for correcting a randomly generated one.
    
\begin{figure}[h]
        \centering
        \subfloat[]{
            \includegraphics[width=0.4\linewidth]{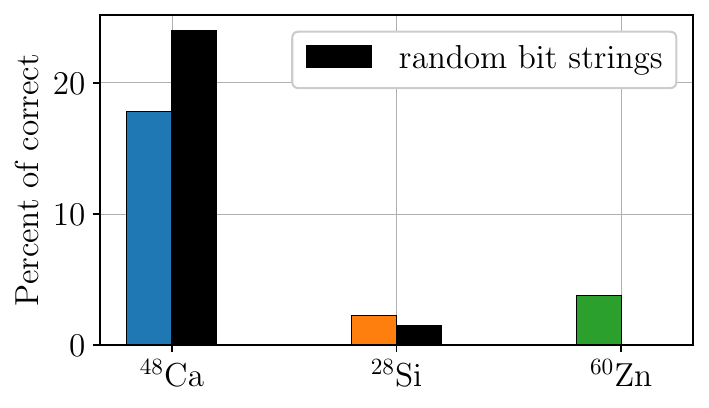}
        }
        \hspace{1.5cm}
        \subfloat[]{
            \includegraphics[width=0.4\linewidth]{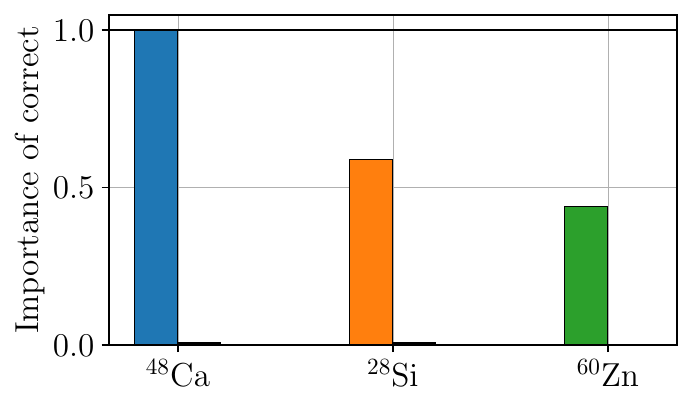}
        }
        \caption{
            (a) Percent of correct sampled bit strings and (b) their importance in the ground state for $^{48}$Ca, $^{28}$Si, and $^{60}$Zn.
            Black bars indicate the data for 250000 bit strings generated randomly and uniformly.
        }
        \label{fig:random_bitstrings-percent-importance}
    \end{figure}

    The contrast is not so stark for $^{48}$Ca and $^{28}$Si.
    Indeed, as one can see in panel (a) of Figure~\ref{fig:random_bitstrings-percent-importance}, sampling a correct bit string (i.e., a bit string with the correct quantum numbers $\nnuc$, $T_z$, $M$ and $\Pi$) from a quantum device is nearly as probable as generating it randomly.
    However, by also looking at panel (b) of this figure, one sees that these random bit strings are much less important for the ground state. 

    The results described above suggest that noisy quantum computers may be used for sampling a number of important bit strings that can serve as a seed for generation of new ones via, e.g., configuration recovery, time-reversal, and excitations.
    In this work, the bit strings are sampled by measuring the Krylov states \eqref{eq:krylov_states} prepared using qDRIFT \eqref{eq:qdrift} with $K=10$ operations. 
    As we discussed in Section~\ref{sec:discussion-operations}, with this rather small number of excitations we can explore only low-excited bit strings.  
    Increasing $K$ may lead to obtaining high-excited bit strings, but also to introducing more noise.
    In Figure~\ref{fig:random_bitstrings-hamming-c}, we show the Hamming distances between the sampled $s \in \mathcal{S}$ and recovered $r \in \mathcal{R}_{\mathcal{S}}$ bit strings for $^{60}$Zn obtained from \texttt{ibm\_basquecountry} with $K=10$ and $K=20$ excitations in the qDRIFT circuit.
    For the larger number of excitations, the distribution of the Hamming distances is considerably closer to that of the randomly generated bit strings.
    Therefore, to explore high-excited bit strings, one needs less noisy hardware.

\end{document}